\documentclass[sigconf]{acmart}

\usepackage{subcaption}
\usepackage{makecell}  %

\newcommand{\revision}[1]{#1}

\AtBeginDocument{%
  }

\copyrightyear{2025}
\acmYear{2025}
\setcopyright{cc}
\setcctype{by}
\acmConference[WWW Companion '25]{Companion Proceedings of the ACM Web Conference 2025}{April 28-May 2, 2025}{Sydney, NSW, Australia}
\acmBooktitle{Companion Proceedings of the ACM Web Conference 2025 (WWW Companion '25), April 28-May 2, 2025, Sydney, NSW, Australia}
\acmDOI{10.1145/3701716.3715543}
\acmISBN{979-8-4007-1331-6/2025/04}

\begin{document}

\title{Wikipedia Contributions in the Wake of ChatGPT}

\author{Liang Lyu}
\email{lianglyu@mit.edu}
\affiliation{%
  \institution{Massachusetts Institute of Technology}
  \city{Cambridge}
  \state{Massachusetts}
  \country{USA}
}

\author{James Siderius}
\email{james.siderius@tuck.dartmouth.edu}
\affiliation{%
  \institution{Tuck School of Business at Dartmouth}
  \city{Hanover}
  \state{New Hampshire}
  \country{USA}
}

\author{Hannah Li}
\email{li.hannahq@gmail.com}
\affiliation{%
  \institution{Columbia Business School}
  \city{New York}
  \state{New York}
  \country{USA}
}

\author{Daron Acemoglu}
\authornote{These authors have contributed equally.}\email{daron@mit.edu}
\affiliation{%
  \institution{Massachusetts Institute of Technology}
  \city{Cambridge}
  \state{Massachusetts}
 \country{USA}
}

\author{Daniel Huttenlocher}
\authornotemark[1]
\email{huttenlocher@mit.edu}
\affiliation{%
  \institution{Massachusetts Institute of Technology}
  \city{Cambridge}
  \state{Massachusetts}
  \country{USA}
}

\author{Asuman Ozdaglar}
\authornotemark[1]
\email{asuman@mit.edu}
\affiliation{%
  \institution{Massachusetts Institute of Technology}
  \city{Cambridge}
  \state{Massachusetts}
  \country{USA}
}

\renewcommand{\shortauthors}{Liang Lyu et al.}

\begin{abstract}
  How has Wikipedia activity changed for articles with content similar to ChatGPT following its introduction?
We estimate the impact using differences-in-differences models, with dissimilar Wikipedia articles as a baseline for comparison, to examine how changes in voluntary knowledge contributions and information-seeking behavior differ by article content. Our analysis reveals that newly created, popular articles whose content overlaps with ChatGPT 3.5 saw a greater decline in editing and viewership after the November 2022 launch of ChatGPT than dissimilar articles did. These findings indicate heterogeneous substitution effects, where users selectively engage less with existing platforms when AI provides comparable content. This points to potential uneven impacts on the future of human-driven online knowledge contributions.
\end{abstract}

\begin{teaserfigure}
  \centering \includegraphics[width=115mm]{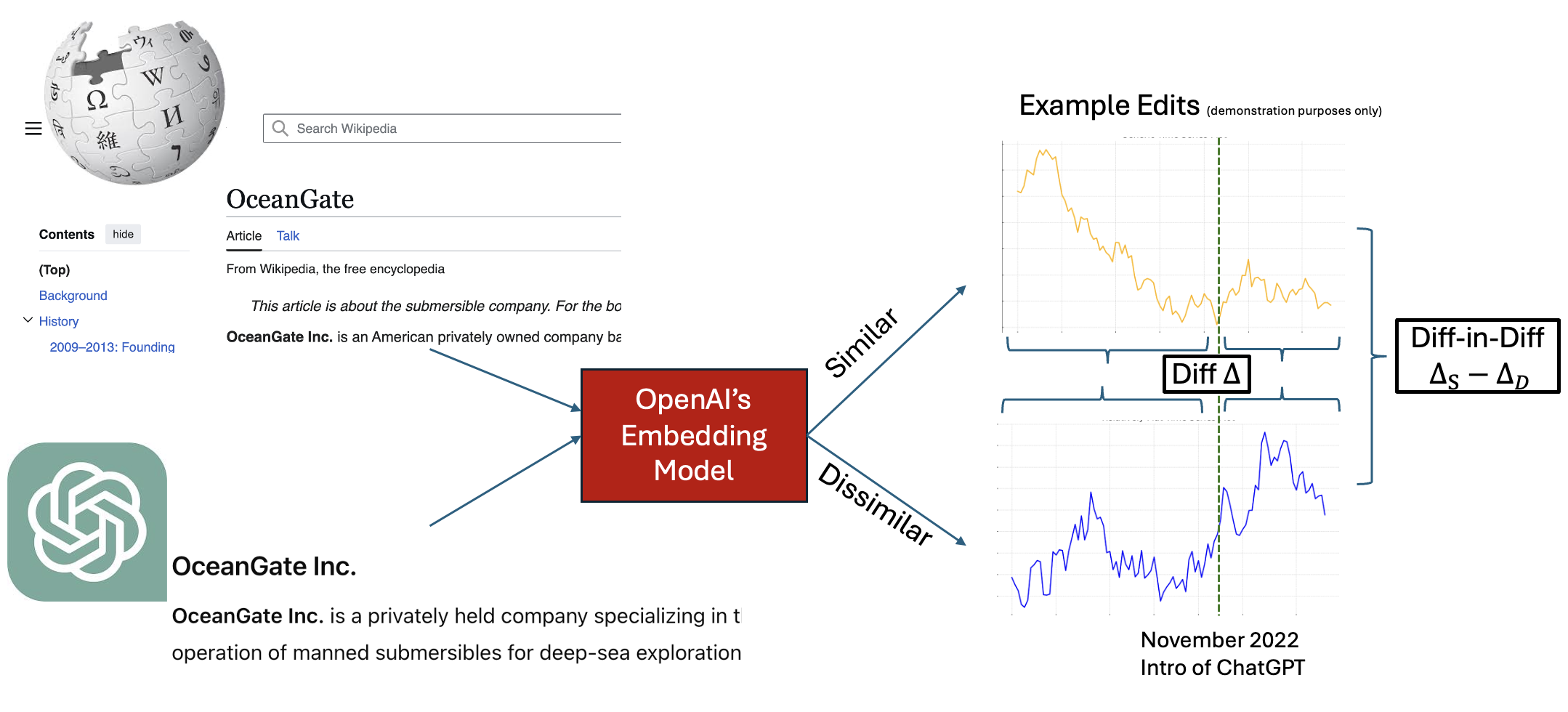}
  \caption{Our Diff-in-Diff Methodology. Measuring Wikipedia activity before and after the introduction of ChatGPT for Wikipedia articles that rank similarly and dissimilarly to the same information generated by ChatGPT.}
  \label{fig:teaser}
\end{teaserfigure}

\maketitle

\section{Introduction}

Generative AI tools like ChatGPT have fundamentally reshaped how information is disseminated and consumed online. Increasingly, users are turning to large language model (LLM) chatbots and AI-generated search summaries for information (\cite{dam2024complete}; \cite{glickman2024ai}). This shift in behavior can create spillover effects on other platforms, as users substitute away from traditional crowdsourced knowledge sources such as Reddit (\cite{burtch2024consequences}) and Stack Overflow (\cite{riochanona2023large}; \cite{shan2023examining}).

Importantly, these changes in user behavior can have downstream impacts not only on these traditional platforms, but also on future generative AI models. One concern is that if users substitute away from these platforms %
and viewership drops, then contributors may also be less inclined to create content and knowledge, which AI models are trained on. 
A decline in their quantity and quality can result in ``model collapse'' for future AI models \cite{shumailov2024ai}. 

In this study, we focus on Wikipedia, a knowledge sharing platform with unique concerns and learning opportunities regarding the potential negative effects of LLMs. Wikipedia's scope extends far beyond technical, directed questions, instead offering an encyclopedic repository of historical, cultural and scientific topics upon which many other sources draw. The platform also hosts two distinct groups of users -- viewers and editors -- who could differ in their responses to LLMs and their downstream implications.

\revision{Several factors about Wikipedia necessitate our differences-in-differences (DiD) strategy, in contrast to the interrupted time series analysis that is often used in similar work on Reddit (\cite{burtch2024consequences}), Stack Overflow (\cite{riochanona2023large}; \cite{shan2023examining}), and Wikipedia itself (\cite{reeves2024exploring}).} In addition to having a broader scope of topics, Wikipedia allows for more diverse user incentives than analogous platforms: viewers exhibit both shallow and deep information needs \cite{singer2017we}, while contributors are driven by \revision{both intrinsic \cite{yang2010motivations} and extrinsic motivations \cite{xu2015empirical}}.
These factors may dampen the effects of ChatGPT on \textit{some} users and articles. In fact, \cite{reeves2024exploring} analyze Wikipedia engagement in the aggregate, and do not identify significant drops in activity following the launch of ChatGPT. We hypothesize that their analysis do not fully capture the heterogeneity of Wikipedia, compared to similar platforms with more homogeneous contents and users.

The heart of our empirical strategy is that the impact of ChatGPT on Wikipedia should vary between different types of articles. We classify a subset of articles as ``similar'' to ChatGPT's output, capturing the capability of the chatbot to learn and summarize relevant information in these articles. A non-overlapping subset of articles are classified as ``dissimilar'', meaning that there is less possibility that ChatGPT will replace Wikipedia as the main provider of information for these articles. We then estimate DiD models with dissimilar articles as our ``control group'', in order to estimate the causal effect of ChatGPT on viewership and edits. 

More specifically, we define our two-year event study period centered at the launch of ChatGPT in November 2022, consider a pool of articles that had been popular before the end of this period and were created relatively recently, and gather their views and edits during the period. We pass their titles to GPT 3.5 and ask it to generate an encyclopedic article, and compute similarity scores between the ChatGPT outputs and the Wikipedia articles. 

This empirical strategy relies on two assumptions.  The first is the ``stable unit treatment value assumption'', which states that our control group, dissimilar articles, is not impacted by ChatGPT. This assumption is motivated by the fact that ChatGPT is less capable of replicating the content of these articles. If the assumption is violated due to indirect effects (e.g., ChatGPT makes Wikipedia contributors more productive), our strategy will still reveal the differential additional effect on similar articles relative to dissimilar ones (but not the full impact on dissimilar articles). Our second assumption is the standard ``parallel trends'' assumption of DiD: absent ChatGPT, similar and dissimilar articles would have evolved in similar fashion. Figure~\ref{fig:time_series_popularnew_single_reg_bootstrap} below, which shows similar pre-ChatGPT evolution of these two groups, bolsters our confidence in this assumption.

Our analysis reveals that, following the launch of ChatGPT, articles with similar content experienced a significant decline in views relative to dissimilar articles. For edits, we find suggestive but not statistically significant evidence of fewer edits on similar articles compared to dissimilar ones. These findings suggest that substitution effects of ChatGPT may vary depending on its capabilities relative to the content of interest. It highlights the need for further research into the potential effects of LLMs on voluntary knowledge contributions across crowdsourced platforms, and mechanisms to explain such behavior for both viewers and content contributors.

\section{Data and Methods}
We perform a difference-in-differences (DiD) analysis of the growth in activities -- views and edits in each month -- of a pool of English Wikipedia pages from the 12-month period preceding ChatGPT's launch to the 12-month period that follows it. Pages are classified as ``similar'' or ``dissimilar'' based on how much ChatGPT's responses on the topic align with the actual Wikipedia pages. We analyze how the growths in activities differ for similar vs. dissimilar articles given fixed recency of articles, controlling for article length and the trend in overall growth of all articles.

\subsection{Wikipedia Data}
\textbf{Time frame.} We define our \textit{event study} period $E$, the period during which activities are measured, as the 24 months from December 2021 to November 2023 (inclusive). It consists of the \textit{pre-GPT} period up to ChatGPT's launch date on November 30, 2022, and the \textit{post-GPT} period starting from December 2022.

\textbf{Articles.} For each month up to November 2023,\footnote{Starting with July 2015, the first month in which viewership data is available} we collect the 1000 pages with the highest number of views on English Wikipedia during that month. 
We obtain 20775 unique articles after data cleaning, denoted as set $\mathcal{A}$.
The 2206 most recently created articles are used in subsequent analyses (since January 2020; see Section~\ref{sec:models}).

The choice of using the most popular articles per month is driven by the motivation to sample articles with substantial levels of activity, whose numbers of views and edits are representative and more robust to noise. Wikimedia APIs provide convenient access to top articles by monthly page views \cite{wiki2025pageviews}.

\textbf{Measurements.} For each article in $\mathcal{A}$, we measure its \textit{activities}, in terms of views and edits, in each month of the event study period. 
An \textit{observation} is defined as $(a, t, Views_{at}, Edits_{at})$, where $a\in \mathcal{A}$ is an article, $t\in E$ is a month in the event study period, and $Views_{at}$ and $Edits_{at}$ are $a$'s number of views and edits in month $t$. 
We also collect the creation date and length of each article.

\subsection{GPT Data and Similarity Score}  \label{sec:gpt_sim}
For each article, we use GPT 3.5 Turbo \cite{gpt2024models} to generate an encyclopedic page given the title. %
This model, trained using data up to September 2021, was the primary model available to most ChatGPT users in 2023, and affected user decisions during the event study period more directly than later models. 

We use the following prompt:
\medskip

\textit{You are an assistant whose task is to write an encyclopedic article for a given topic chosen by the user, similar to those found on Wikipedia.} %

\textit{Generate an encyclopedic article in English with title "\textit{[title of actual Wikipedia page]}".
}
\medskip

We compute a \textit{similarity score} for each article between its actual contents on Wikipedia and GPT 3.5's imitation, as the cosine similarity of their text embeddings using OpenAI's embedding model \texttt{text-embedding-3-small}. \cite{gpt2024embedding} The similarity score can be seen as a proxy for substitutability of the two options from the user's point of view, and for GPT 3.5's mastery of the topic.

\subsection{Models}  \label{sec:models}
\newcommand{\simbinary}{s_T(a)}%
\textbf{Recency}. \revision{We parameterize the \textit{recency} of articles in our sample by parameter $T$, and perform a series of regressions by varying $T$. For each $T$, an observation $(a, t, Views_{at}, Edits_{at})$ is called \textit{$T$-recent} at time $t$ if the article $a$ was created no more than $T$ months before~$t$. We define $\mathcal{O}(t,T)$ as the set of all $T$-recent observations recorded in month $t$, and $\mathcal{O}(T) = \bigcup_{t\in E} \mathcal{O}(t,T)$ the entire set of $T$-recency observations for some $t$ in the event study period.}

Varying the recency parameter $T$ lets us analyze activities of articles of different ages. A small $T$ means that we only look at articles created during or just before the event study period, and their activities soon after they were created. A larger $T$ shifts the focus to older articles and their activities further away from creation.

\textbf{Similarity classifications}. We then compute similarity labels for each regression parameterized by $T$. Define $\mathcal{A}(T)$ as the set of all unique articles that are involved in observations $\mathcal{O}(T)$.
Each article $a\in \mathcal{A}(T)$ is then given a binary classification label, $\simbinary{} \in \{ 0,1 \}$, such that $\simbinary{} = 1$ iff its similarity score is above the median in $\mathcal{A}(T)$. These are considered ``similar'' articles, whereas those with $\simbinary{} = 0$ are ``dissimilar'' articles. We apply 5\% winsorization to views and edits (separately) of articles in $\mathcal{A}(T)$ to remove outliers.

\textbf{Comparative time series.} To analyze how activities of similar and dissimilar articles change over time, we fit the following regression model for a fixed $T$, with data as all observations in $\mathcal{O}(t,T)$ for all months $t\in E$:
\begin{align*}
    Views_{at} &= \delta_l \, length(a) + \sum_{t'} \theta_{t'}{\bf 1}_{CreationMonth(a) = t'} + \delta_t t + \epsilon_{at},
\end{align*}
where $\delta_l$ and $\delta_t$ are linear trends for article length and activity time. An analogous model is defined for edits. We report the time series of mean residuals for each month of activity $t$ aggregated over similar and dissimilar articles respectively, with bootstrapped standard errors. The results for $T=6$ are presented in Section~\ref{sec:results_ts}.

\textbf{Difference-in-differences (DiD) analysis.} 
We fit a regression model for each $T$ to estimate the DiD using the same data as above:
\begin{align*}
    Views_{at} &= \gamma_1 \simbinary{} + \gamma_2 \times {\bf 1}_{t \geq \text{Dec 2022}} + \beta \left( \simbinary{} \times {\bf 1}_{t \geq \text{Dec 2022}} \right) \\
    &\phantom{{}={}} + \delta_l \, length(a) + \sum_{t'} \theta_{t'} {\bf 1}_{CreationMonth(a)=t'} + \delta_t t + \epsilon_{at}.
\end{align*}
The coefficient of interest is $\beta$, which measures the effect of ChatGPT on similar articles. $\gamma_1$, $\gamma_2$ and $\theta$ are fixed effects for similarity, activity occurring post-GPT, and creation month of the article, respectively. $\delta_l$ and $\delta_t$ are linear trends for article length and activity time.
An analogous model is defined for edits.

To smooth the estimates for various $T$, we perform weighted least squares (WLS) where each observation $(a,t)$ has weight $\alpha^{T-\text{Age}_{at}}$, where $\text{Age}_{at}$ is the age of article $a$ at time $t$, for a chosen smoothing parameter $\alpha$. The results are presented in Section~\ref{sec:results_did}.

\section{Results}
\label{sec:main_results}

\subsection{Comparative Time Series} \label{sec:results_ts}

\begin{figure}[h]
    \centering
    \begin{subfigure}[t]{0.4\textwidth}
        \centering
        \includegraphics[width=\linewidth]{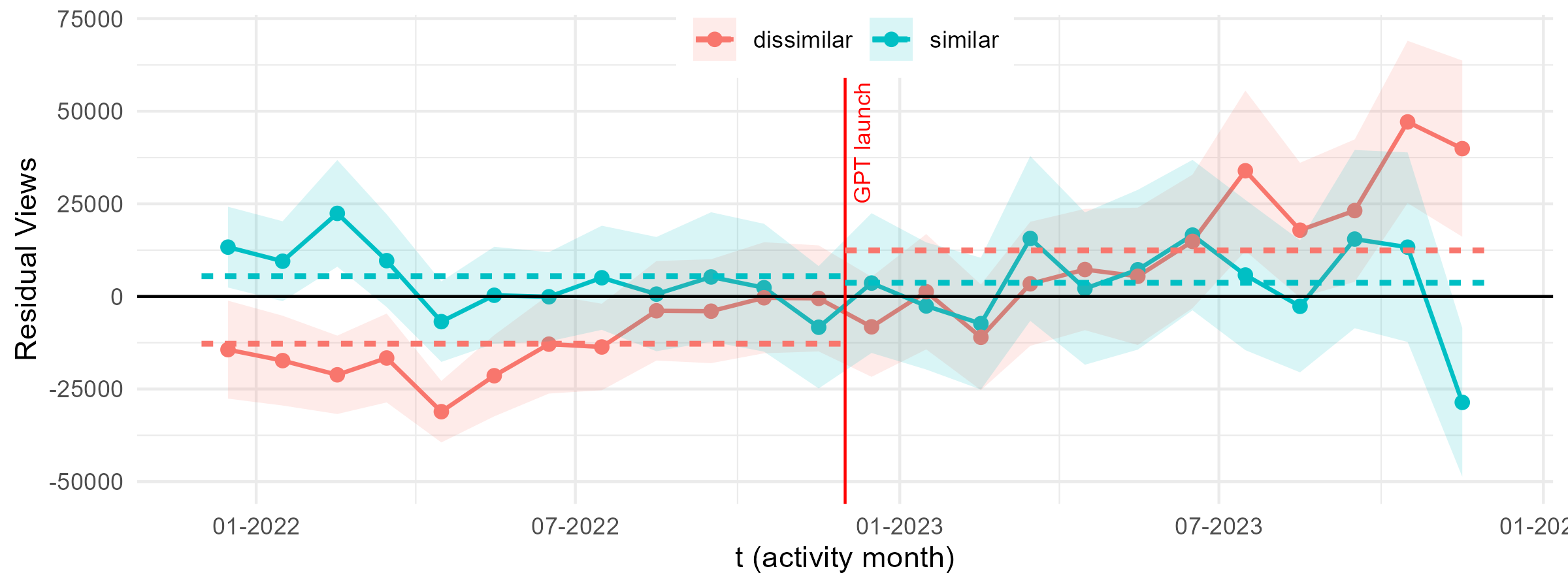}
        \caption{Views}
    \end{subfigure}
    \hfill
    \begin{subfigure}[t]{0.4\textwidth}
        \centering
        \includegraphics[width=\linewidth]{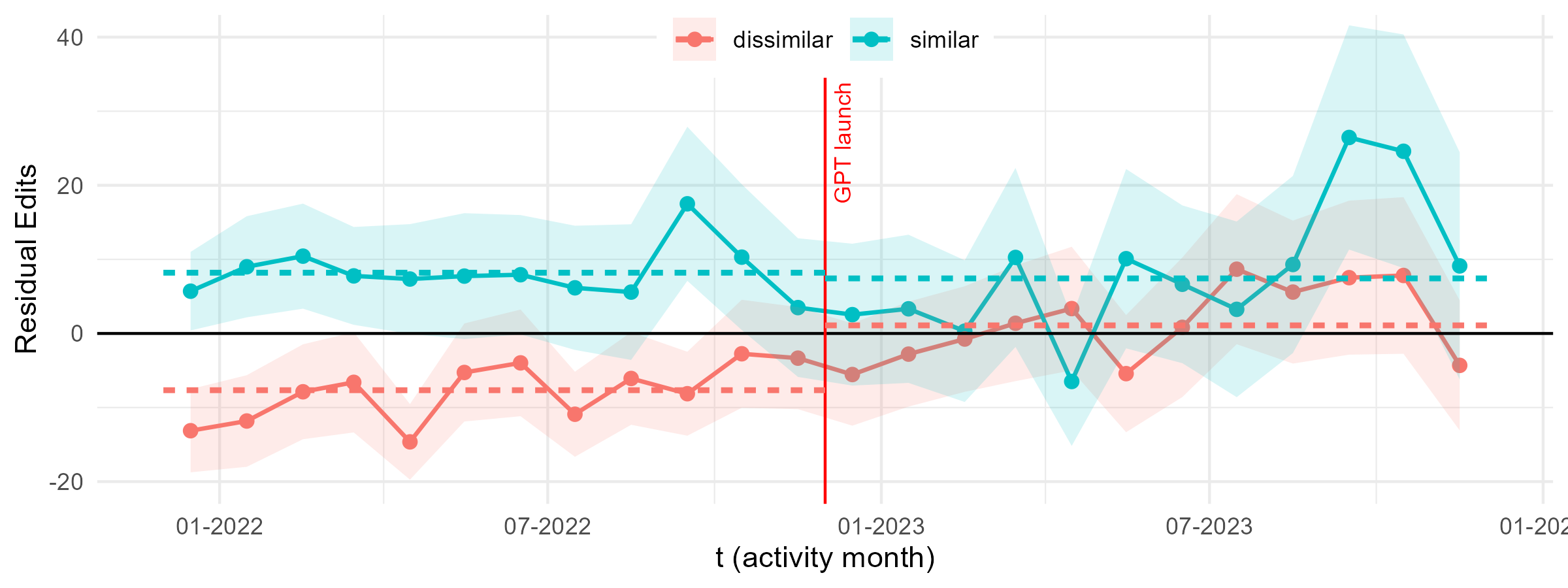}
        \caption{Edits}
    \end{subfigure}
    \caption{Mean residuals for each month $t$ in which activity occurs, among all observations in $\mathcal{O}(6)$ (at most 6 months old), with bootstrapped standard errors with $1000$ samples. Dashed lines are mean residuals for similar (blue) and dissimilar (red) articles over the pre-GPT and post-GPT periods respectively. Red vertical line indicates date of ChatGPT launch.}
    \label{fig:time_series_popularnew_single_reg_bootstrap}
\end{figure}

Figure~\ref{fig:time_series_popularnew_single_reg_bootstrap} shows a comparative time series of mean residual views and edits in each month during the event study period for articles that are at most $T=6$ months old.
For both views and edits, we see that similar articles exhibited little changes in activity from the pre-GPT to the post-GPT period (accounting for controls). Dissimilar articles, on the other hand, show an increase in edits after ChatGPT launched in November 2022 (Figure~\ref{fig:time_series_popularnew_single_reg_bootstrap}(b)). Views for dissimilar articles also show an increasing trend around the same time, with residuals even rising above those of similar articles (Figure~\ref{fig:time_series_popularnew_single_reg_bootstrap}(a)).

\subsection{Diff-in-Diff Regressions} \label{sec:results_did}
\begin{figure}[htbp]  %
    \centering
    \begin{subfigure}[t]{0.4\textwidth}
        \centering
        \includegraphics[width=\linewidth]{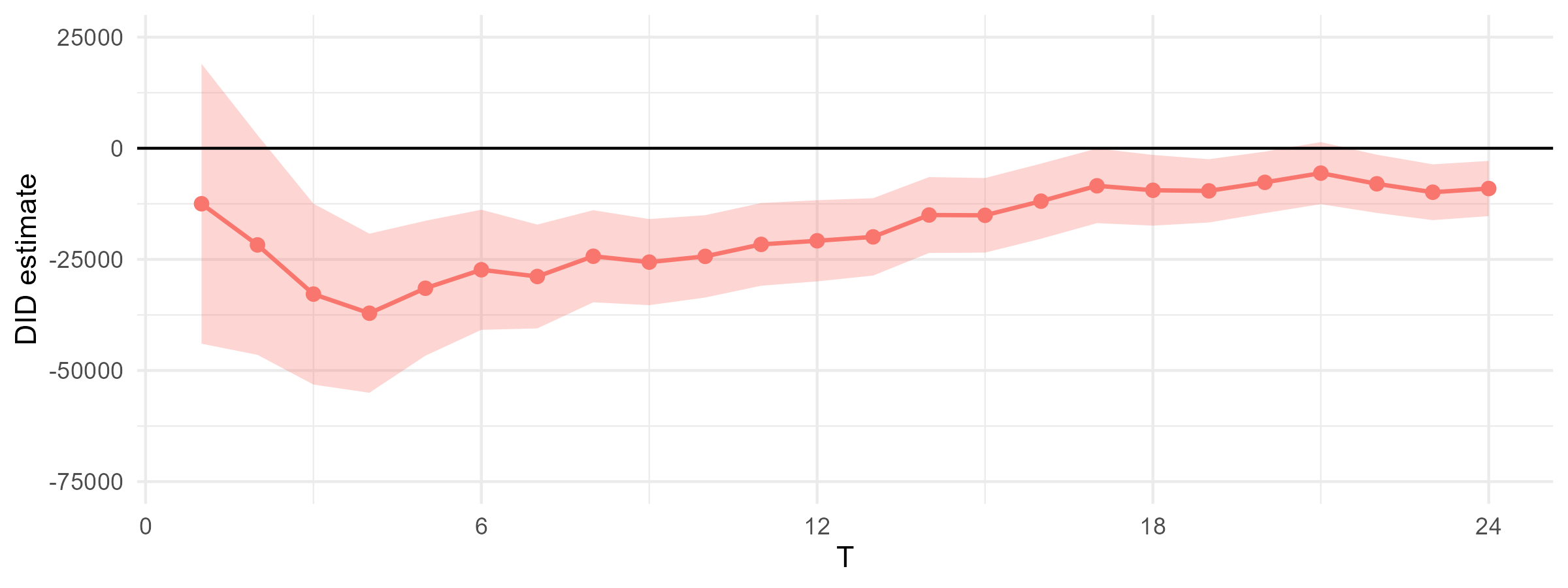}
        \caption{Views}
    \end{subfigure}
    \hfill
    \begin{subfigure}[t]{0.4\textwidth}
        \centering
        \includegraphics[width=\linewidth]{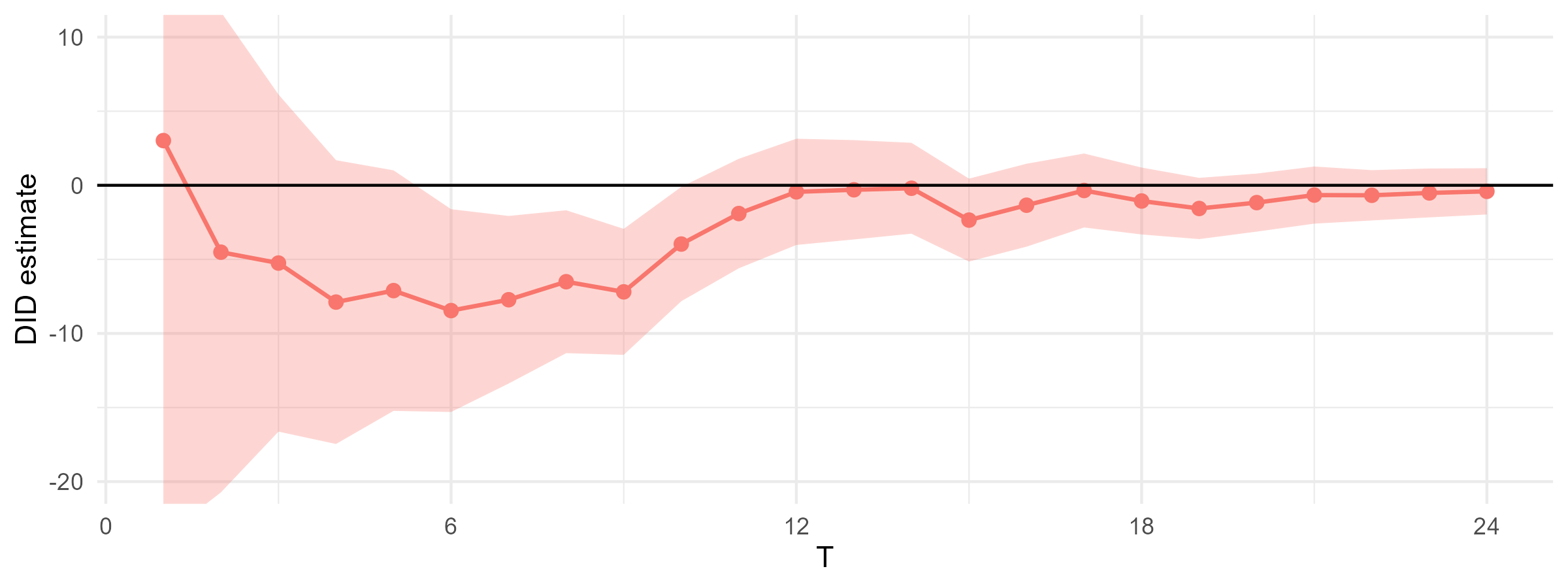}
        \caption{Edits}
    \end{subfigure}
    \caption{
    Estimated DiD coefficients and 95\% confidence intervals (HC1) for recency parameters $T\in \{1, \dots, 24\}$ on the $x$-axis, with smoothing factor $\alpha=0.8$. Each point is from a regression using observations in $\mathcal{O}(T)$. Observations when the article is closer to $T$ months old given greater weight.}
    \label{fig:did_3.5small_popularnewA}
\end{figure}

Figure~\ref{fig:did_3.5small_popularnewA} shows the estimated diff-in-diff coefficients for each recency parameter $T$ from $1$ to $24$ months ($x$-axis), with smoothing factor $\alpha=0.8$. 
\revision{Each point represents an individual regression: points to the left focus on newer articles and their views and edits in the first few months, while points to the right place greater weight on activities of older articles close to two years after creation.}
Table~\ref{table:reg} shows the estimates for activities that occurred within at most $T=6$ months since article creation, with $\alpha=1$.

\newlength{\rowspace}
\setlength{\rowspace}{-2.4ex}

\begin{table}[htbp] \centering   
  \caption{Diff-in-diff regression of views and edits with recency $T=6$ (activities within 6 months of article creation), with $\alpha=1$ (all articles given equal weight).  } 
  \label{table:reg} 
\begin{tabular}{lcc}   %
\\[\rowspace]%
\hline 
\hline \\[\rowspace] 
\\[\rowspace] & Views & Edits  \\ 
\hline \\[\rowspace] 
 Diff-in-Diff Estimate & $-$29,013.340$^{***}$ & $-$6.877$^{*}$ \\ 
  & (6,903.964) & (3.719) \\ 
  & \\ [\rowspace] 
\hline \\[\rowspace] 
Observations & 6,060 & 6,060 \\ 
R$^{2}$ & 0.069 & 0.140 \\ 
Adjusted R$^{2}$ & 0.064 & 0.135 \\ 
\hline 
\hline \\[\rowspace] 
\textit{Note:}  & \multicolumn{2}{r}{$^{*}$p$<$0.1; $^{**}$p$<$0.05; $^{***}$p$<$0.01} \\ [\rowspace] 
\end{tabular} 
\end{table}

The diff-in-diff coefficients for Figure \ref{fig:did_3.5small_popularnewA}a (views) are negative and statistically significant for all article ages except $T=1$, which implies that Wikipedia articles where ChatGPT provides a similar output experience a larger drop in views after the launch of ChatGPT. 
This effect is much less pronounced for edit behavior. The coefficients for Figure \ref{fig:did_3.5small_popularnewA}b (edits) are generally close to zero and not statistically significant, particularly for articles older than 10 months. For newer articles less than 10 months old, the results are suggestive, but not necessarily significant, that similar articles receive fewer edits than dissimilar articles.

\subsection{Interpretation}
Both analyses show that similar articles that are created recently ($T\leq 10$) experience a greater drop in viewership and contributions post-GPT relative to dissimilar ones, with much stronger statistical significance for views. This shows that the impacts of ChatGPT are heterogeneous, and correlate with its substitutability over Wikipedia in providing information on relevant topics.

\revision{We also observe that the drop in views and especially edits of older articles are smaller in magnitude, evident from regressions with greater $T$. When $T>10$, the effects on edits (if any) are no longer statistically significant, unlike views.}

\revision{There are plausible mechanisms that could explain our findings, though an in-depth analysis of this question is beyond the scope of this paper.} 
The majority of Wikipedia viewers have shallow information needs such as a quick factual lookup \cite{singer2017we, lim2009college}, a task that may be better suited to generative AI models. 
\revision{
Contributors, on the other hand, are often driven by both intrinsic incentives \cite{yang2010motivations} and extrinsic factors such as reciprocity \cite{xu2015empirical}. Some editors, but not all, may thus become less motivated following declines in viewership.
}

\section{Conclusion and Future Work}

This work shows evidence of heterogeneous substitution patterns in Wikipedia viewership. The diff-in-diff analysis shows that articles that are similar to ChatGPT generated content receive fewer views after the launch of ChatGPT, compared to dissimilar articles. 
We also identify a suggestive, though inconclusive, trend in which these similar articles output receive fewer edits, particularly new articles.

The findings on edit activity could have important implications for AI models and warrant further study. A reduction in contributions on Wikipedia could degrade the quality of future training data. 
This phenomenon should be monitored over time, since contributor behavior may react more slowly than reader behavior. 
\revision{
Other future directions include behavioral studies to explain the mechanisms behind the observed trends, and refinements in measuring substitutability of LLMs over Wikipedia as new models become more capable of learning and replicating Wikipedia content.}

\bibliographystyle{ACM-Reference-Format}
\bibliography{bibliography}

\newif\ifShowAppendix
\ShowAppendixfalse

\ifShowAppendix

\appendix

\section{Examples of Similar and Dissimilar Articles}

\newcommand{\mysize}{\footnotesize}
\newcommand{\pagecontent}[1]{{\mysize #1}}
\font\tenipa=tipa10
\def\schwa{{\tenipa\char64}}

\begin{table*}
  \caption{Examples of similar and dissimilar articles}
  \label{table:article_examples}
  \begin{tabular}{|p{2.5cm}|p{2cm}|p{6cm}|p{6cm}|}
    \hline
    Similarity and Possible Reasons & Title & Wikipedia & GPT 3.5 \\
    \hline
    Similar (Well-established concept) & Waffle House Index  & \pagecontent{The Waffle House Index is a metric named after the ubiquitous Southern US restaurant chain Waffle House known for its 24-hour, 365-day service. Since this restaurant always remains open, it has given rise to an informal but useful metric to determine the severity of a storm and the likely scale of assistance required for disaster recovery. It was coined by former administrator Craig Fugate of the Federal Emergency Management Agency (FEMA). The metric is unofficially used by FEMA to inform disaster response.} & \pagecontent{The Waffle House Index is an informal metric used by the Federal Emergency Management Agency (FEMA) to determine the impact of a natural disaster in the United States. Named after the American restaurant chain Waffle House, the index was coined by former FEMA administrator Craig Fugate in 2004 following Hurricane Charley. The index is based on the extent of operations at Waffle House restaurants in areas affected by disasters, particularly hurricanes and tornadoes.}  \\
    \hline
    Similar (Recent information that GPT has been trained on) & The Lost Daughter (film) & \pagecontent{The Lost Daughter is a 2021 psychological drama film written and directed by Maggie Gyllenhaal (in her feature directorial debut), based on the 2006 novel of the same name by Elena Ferrante. The film stars Olivia Colman, Dakota Johnson, Jessie Buckley, Paul Mescal, Dagmara Domińczyk, Jack Farthing, Oliver Jackson-Cohen, Peter Sarsgaard, and Ed Harris. Colman also served as an executive producer.} & \pagecontent{"The Lost Daughter" is a drama film written and directed by Maggie Gyllenhaal. The film is based on the novel of the same name by Elena Ferrante and premiered at the 2021 Venice Film Festival. It was later released on Netflix in December 2021.}  \\
    \hline
    Dissimilar (Lack of up-to-date information) & Israel--Hamas war & \pagecontent{An armed conflict between Israel and Hamas-led Palestinian militant groups and their allies has been taking place chiefly in the Gaza Strip with other confrontations in the Gaza Envelope and southern Israel, at the Israel-Lebanon border and in the West Bank since October 2023. 
    } & \pagecontent{The Israel--Hamas conflict refers to the long-standing and ongoing hostilities between the State of Israel and the Palestinian militant group Hamas. This protracted conflict is rooted in the complex history and competing national aspirations of both parties, as well as the broader Israeli-Palestinian conflict.}  \\
    \hline
    Dissimilar (Hallucination) & George Santos & \pagecontent{George Anthony Devolder Santos (born July 22, 1988) is an American politician who served as the U.S. representative for New York's 3rd congressional district from January to December 2023, before he was expelled from Congress.} & \pagecontent{George Santos is a renowned contemporary artist known for his unique abstract paintings and sculptures. Born on June 12, 1978, in New York City, Santos discovered his passion for art at a young age and pursued formal training in fine arts at the prestigious School of Visual Arts.}  \\
    \hline
    Dissimilar (Lack of new information, wrong entity) & Barbenheimer & \pagecontent{Barbenheimer ( BAR-b\schwa n-hy-m\schwa r) was a cultural phenomenon which preceded and surrounded the simultaneous theatrical release of two films, Warner Bros. Pictures' Barbie and Universal Pictures' Oppenheimer, on July 21, 2023. 
    } & \pagecontent{Barbenheimer is a small rural village located in the Rhineland-Palatinate region of Germany. Situated in the beautiful countryside near the city of Mainz, Barbenheimer is known for its picturesque surroundings, historic charm, and vibrant local community.}  \\
    \hline
    Dissimilar (Different levels of details) & Project Y & \pagecontent{The Los Alamos Laboratory, also known as Project Y, was a secret laboratory established by the Manhattan Project and operated by the University of California during World War II. Its mission was to design and build the first atomic bombs. Robert Oppenheimer was its first director, serving from 1943 to December 1945, when he was succeeded by Norris Bradbury.} & \pagecontent{Project Y is a highly secretive research initiative that is shrouded in mystery and speculation. The details and goals of the project are largely unknown to the public, as it is classified at the highest levels of security. Project Y has captivated the imagination of conspiracy theorists and enthusiasts of science fiction alike, leading to numerous theories and rumors surrounding its purpose and potential implications.}  \\
    \hline
  \end{tabular}
\end{table*}

Table~\ref{table:article_examples} shows a list of representative Wikipedia articles and corresponding GPT 3.5 responses (generated from April to June 2024 given the article title). They are chosen among articles with the highest (similar) and lowest (dissimilar) similarity scores in the dataset.

\section{Results for Older Articles}

Figure~\ref{fig:did_3.5small_popularnewA} in the main text presents results of diff-in-diff coefficients for recency windows $T$ from 1 to 24 months. Figure~\ref{fig:appx_T60} extends the results to articles that were older during the event study period, up to a maximum of 60 months old when their views and edits were recorded. 
The results largely agree with our findings in the main text, showing a statistically significant effect that similar articles experience a greater decline in viewership after the launch of ChatGPT compared to dissimilar articles, as well as suggestive but less pronounced effects for edits. 

\begin{figure}[htbp]
    \centering
    \begin{subfigure}[t]{0.4\textwidth}
        \centering
        \includegraphics[width=\linewidth]{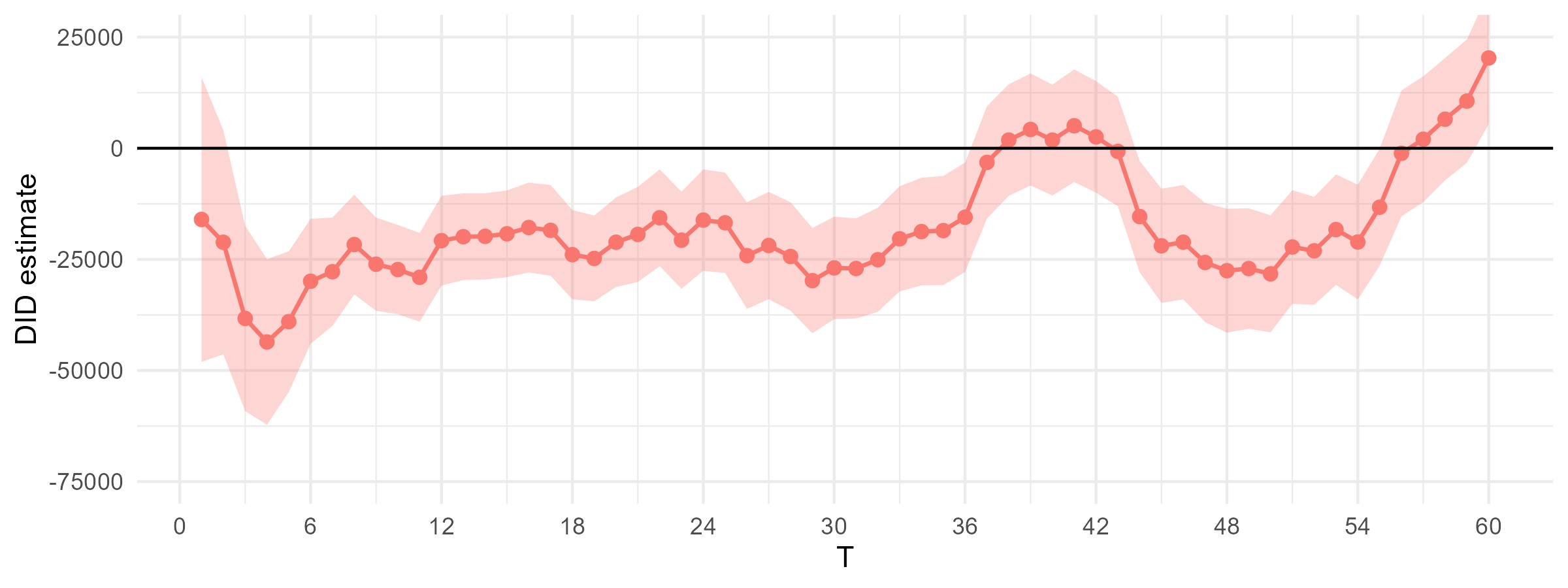}
        \caption{Views}
    \end{subfigure}
    \hfill
    \begin{subfigure}[t]{0.4\textwidth}
        \centering
        \includegraphics[width=\linewidth]{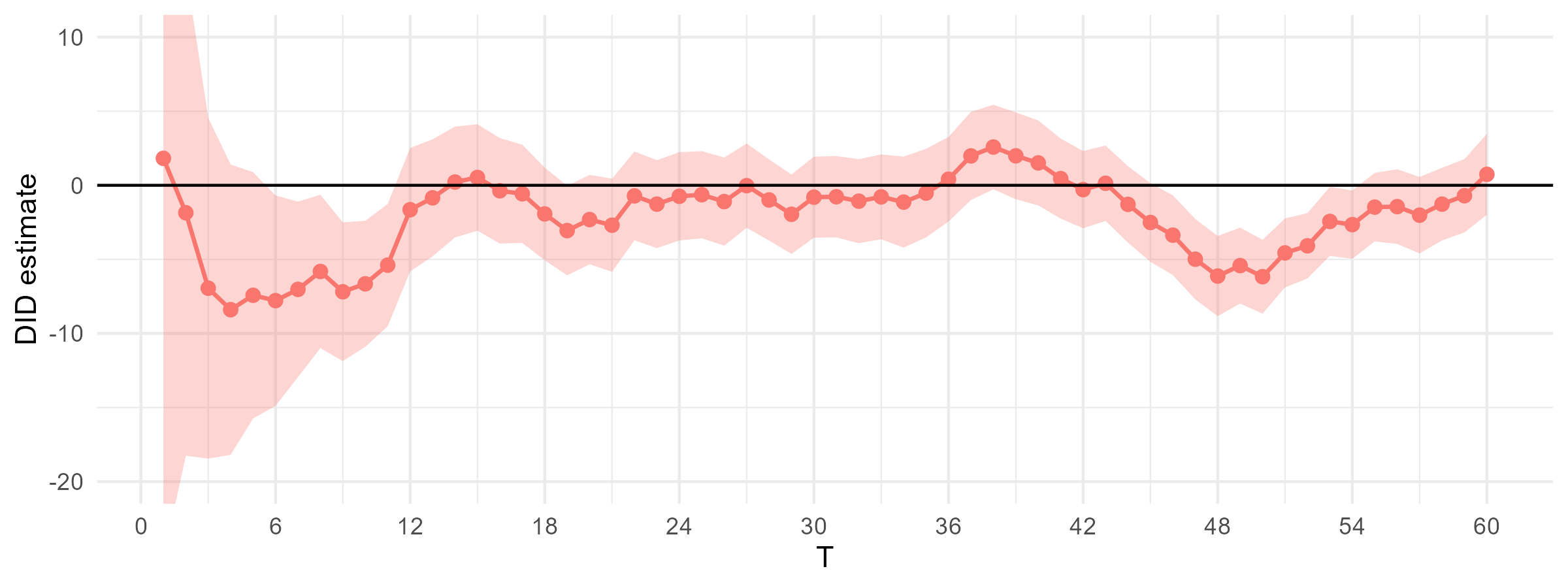}
        \caption{Edits}
    \end{subfigure}
    \caption{
    Estimated DiD coefficients and 95\% confidence intervals (HC1) for each recency window $T\in \{1, \dots, 60\}$, with smoothing factor $\alpha=0.8$.
    }
    \label{fig:appx_T60}
\end{figure}

One notable difference is that for articles of certain ages, namely around 40 months and 60 months, the diff-in-diff estimator is no longer negative nor statistically significant. A possible reason is that ChatGPT was already trained on such articles, as they were published before September 2021 (date of GPT 3.5's training data) and had reasonable popularity. This factor will be discussed further in Appendix~\ref{sec:appx_gpt4}.

\section{Robustness Checks}
In this section, we present variations of parameters of our model. We show that they do not fundamentally affect our main results from the comparative time series and diff-in-diff regressions: that similar articles experience a greater drop in views (statistically significant) and edits (suggestive but not statistically significant) than dissimilar articles.

\subsection{Exponential Smoothing Factors}
In the main text, we applied an exponential smoothing factor of $\alpha=0.8$ to the diff-in-diff weighted least squares (WLS) regression for each recency window $T$, such that articles whose ages are closer to $T$ are given greater weight.
Figure~\ref{fig:appx_expsmooth} shows results using different smoothing factors.

\begin{figure}[h]
    \centering
    \begin{subfigure}[t]{0.22\textwidth}
        \centering
        \includegraphics[width=\linewidth]{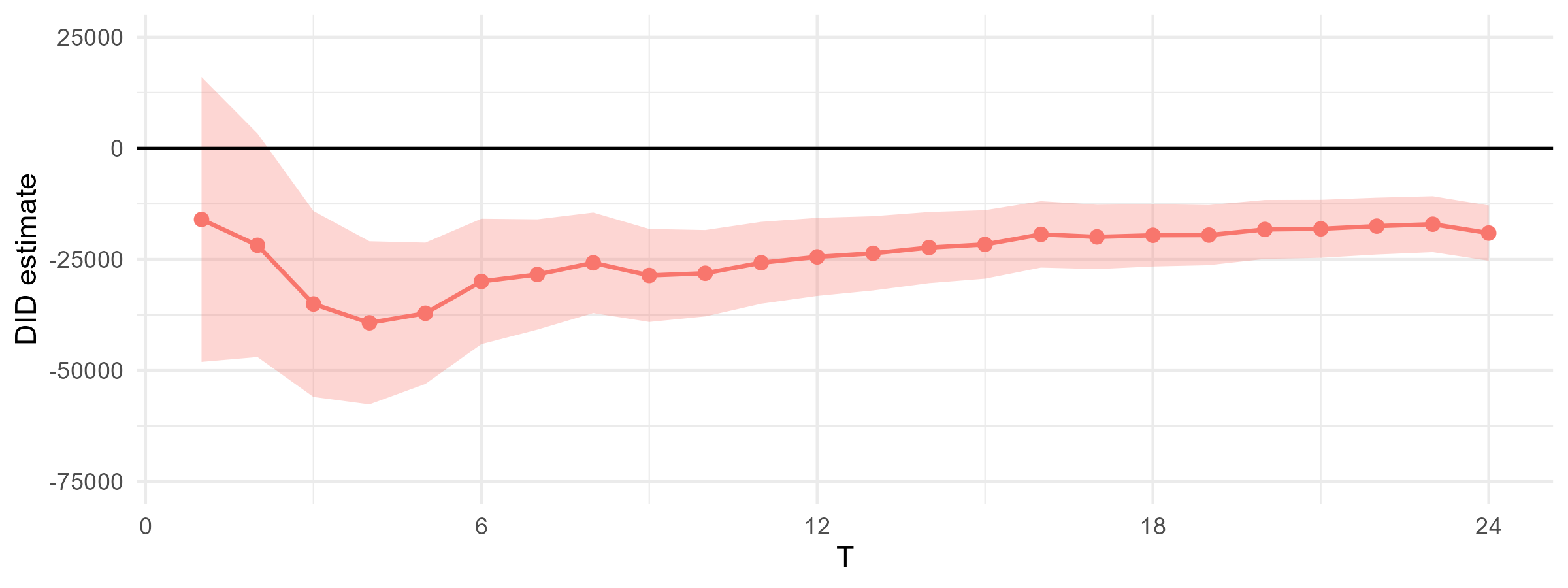}
        \caption{Views, $\alpha=1$}
    \end{subfigure}
    \hfill
    \begin{subfigure}[t]{0.22\textwidth}
        \centering
        \includegraphics[width=\linewidth]{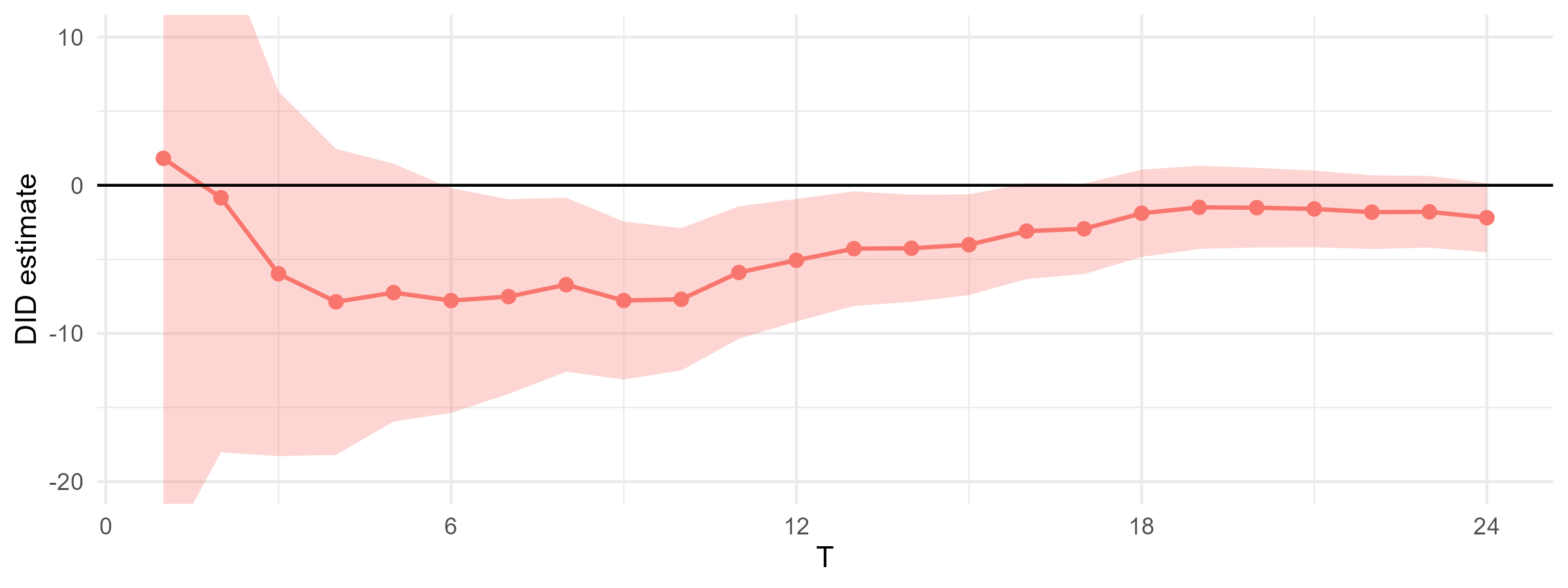}
        \caption{Edits, $\alpha=1$}
    \end{subfigure}
    \hfill
    \begin{subfigure}[t]{0.22\textwidth}
        \centering
        \includegraphics[width=\linewidth]{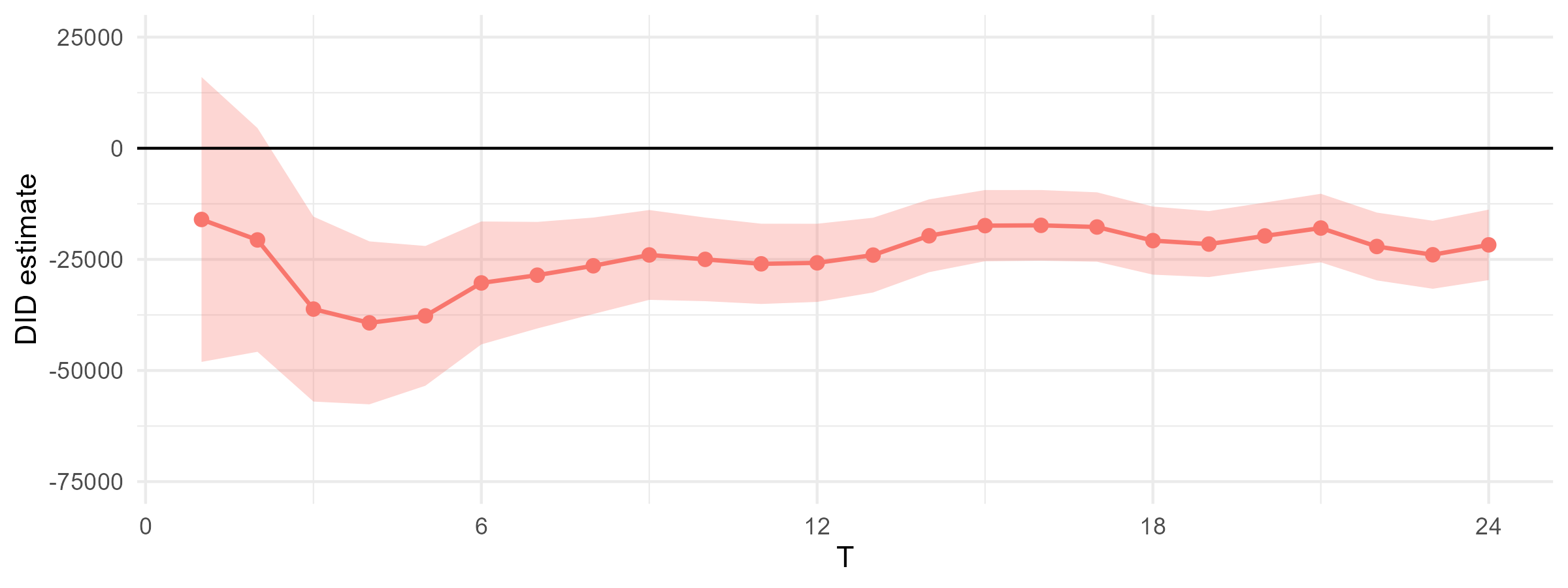}
        \caption{Views, $\alpha=0.9$}
    \end{subfigure}
    \hfill
    \begin{subfigure}[t]{0.22\textwidth}
        \centering
        \includegraphics[width=\linewidth]{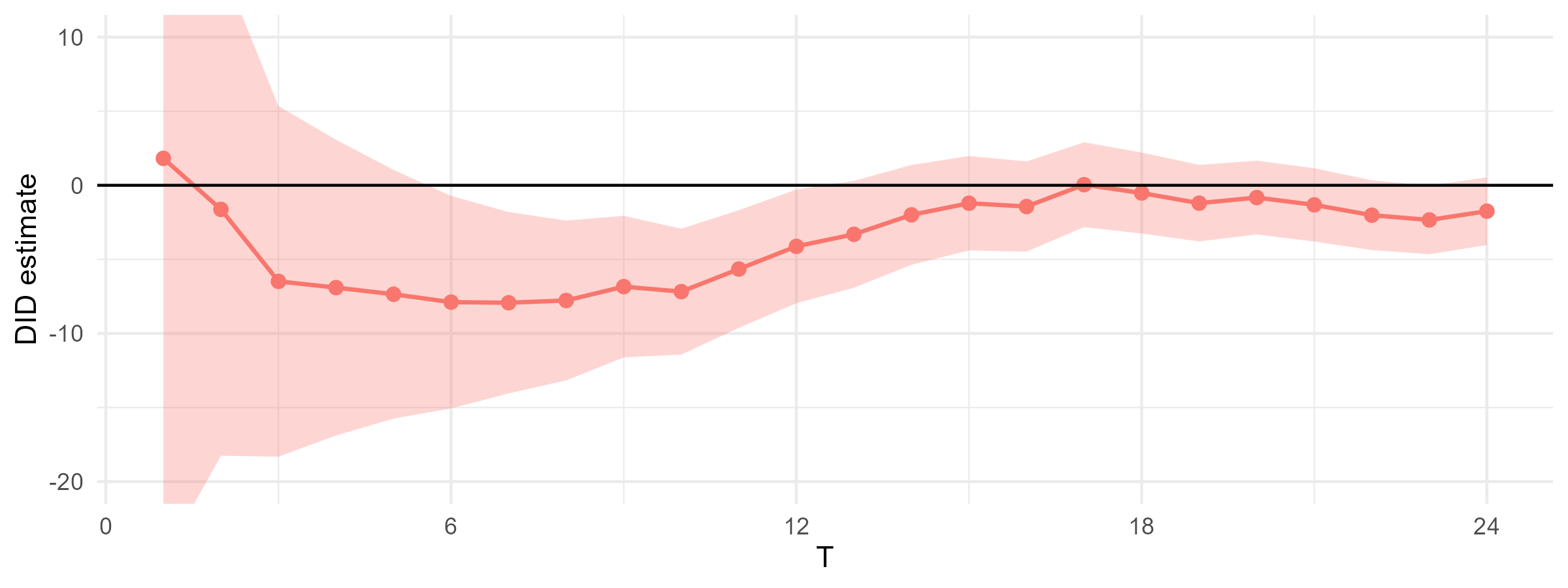}
        \caption{Edits, $\alpha=0.9$}
    \end{subfigure}
    \hfill
    \begin{subfigure}[t]{0.22\textwidth}
        \centering
        \includegraphics[width=\linewidth]{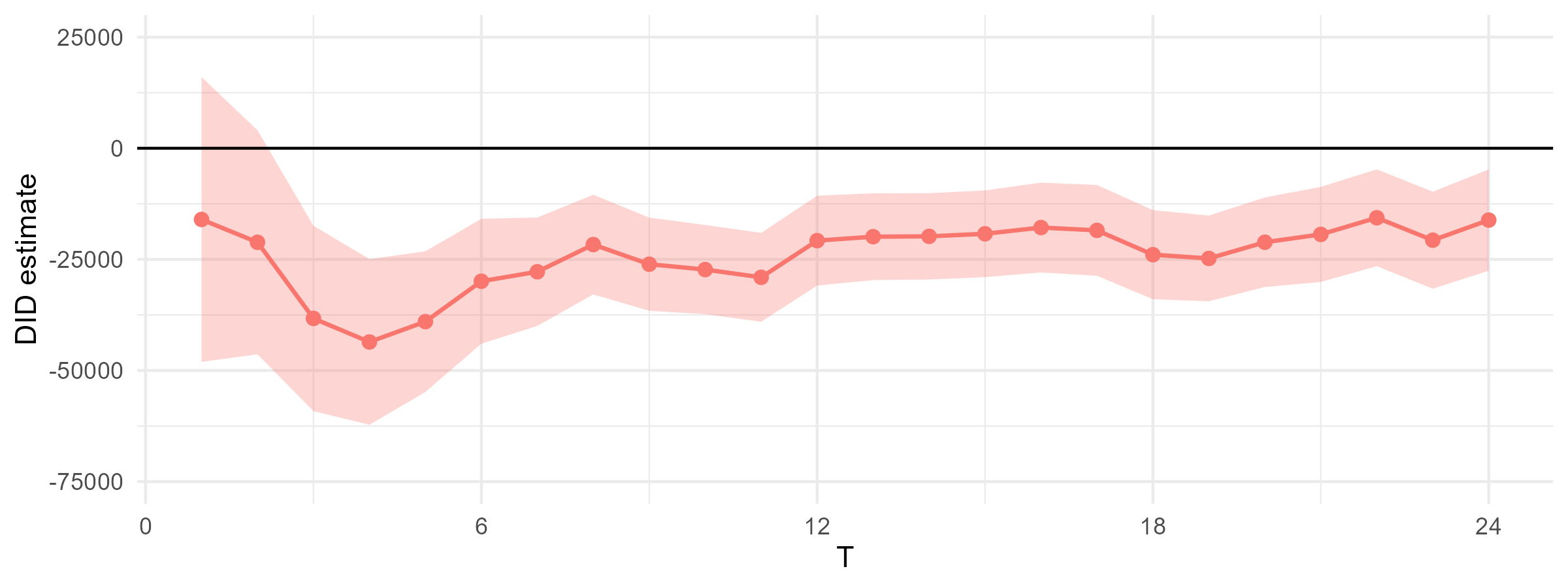}
        \caption{Views, $\alpha=0.8$}
    \end{subfigure}
    \hfill
    \begin{subfigure}[t]{0.22\textwidth}
        \centering
        \includegraphics[width=\linewidth]{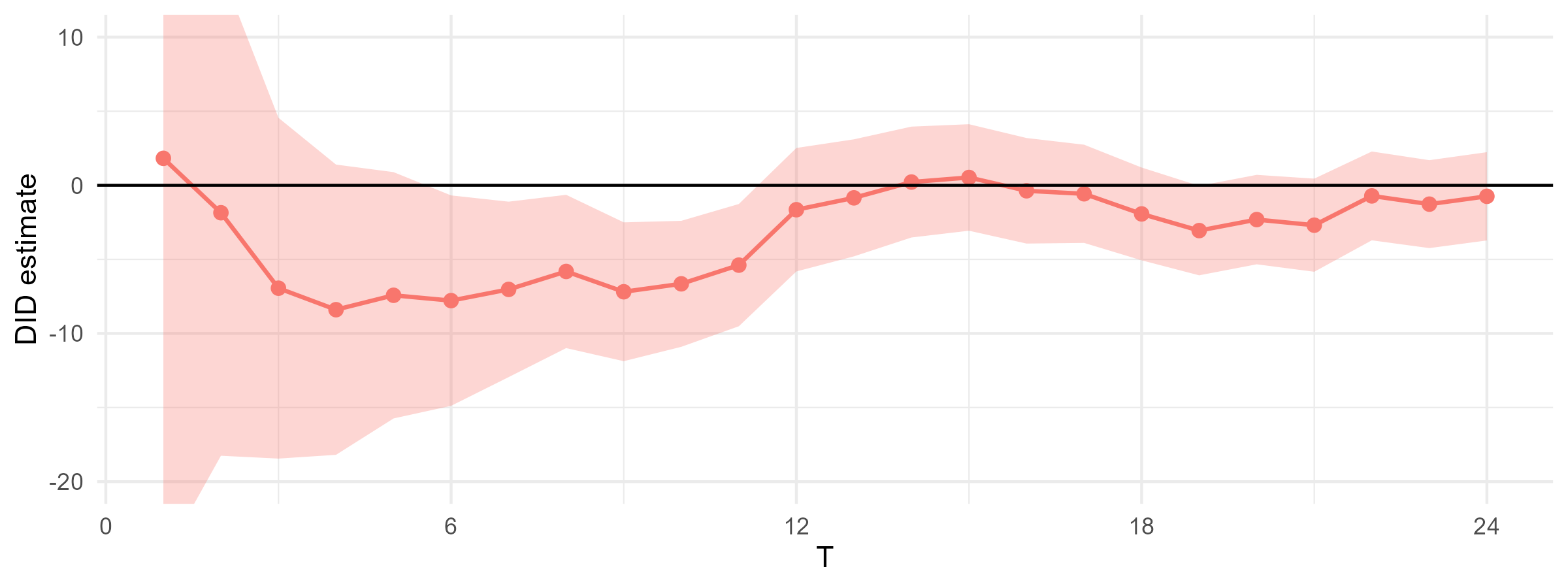}
        \caption{Edits, $\alpha=0.8$}
    \end{subfigure}
    \hfill
    \begin{subfigure}[t]{0.22\textwidth}
        \centering
        \includegraphics[width=\linewidth]{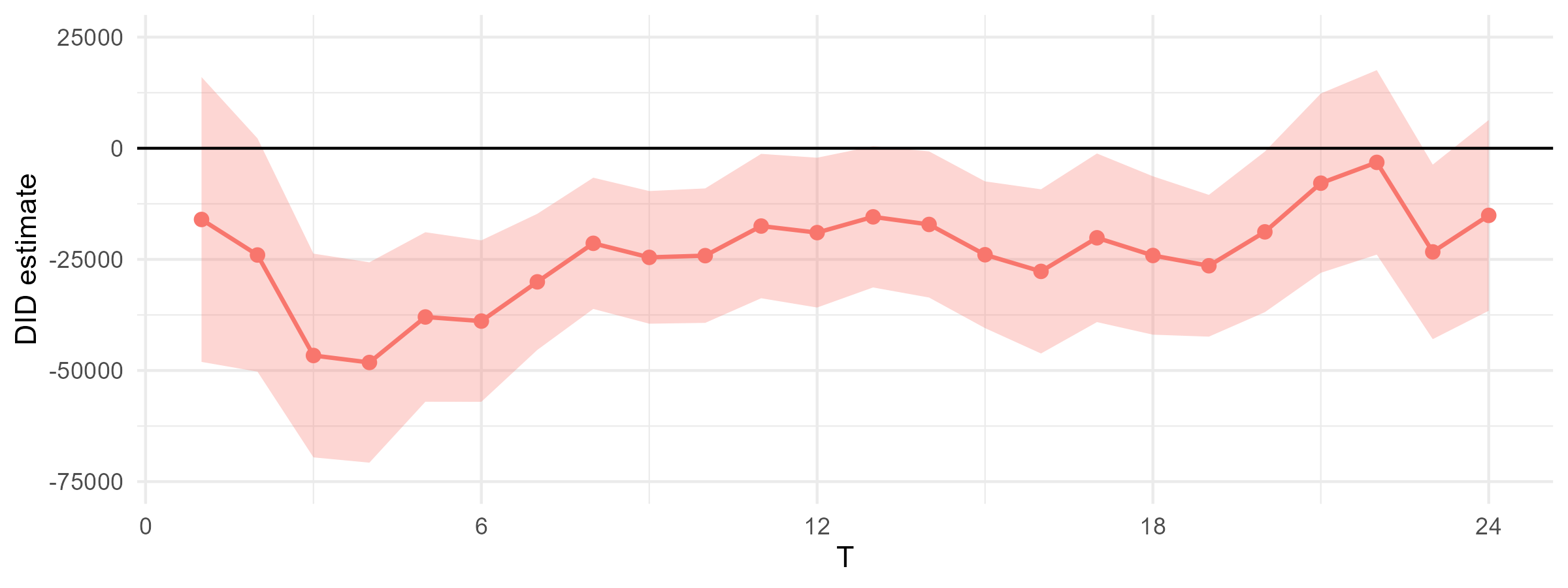}
        \caption{Views, $\alpha=0.5$}
    \end{subfigure}
    \hfill
    \begin{subfigure}[t]{0.22\textwidth}
        \centering
        \includegraphics[width=\linewidth]{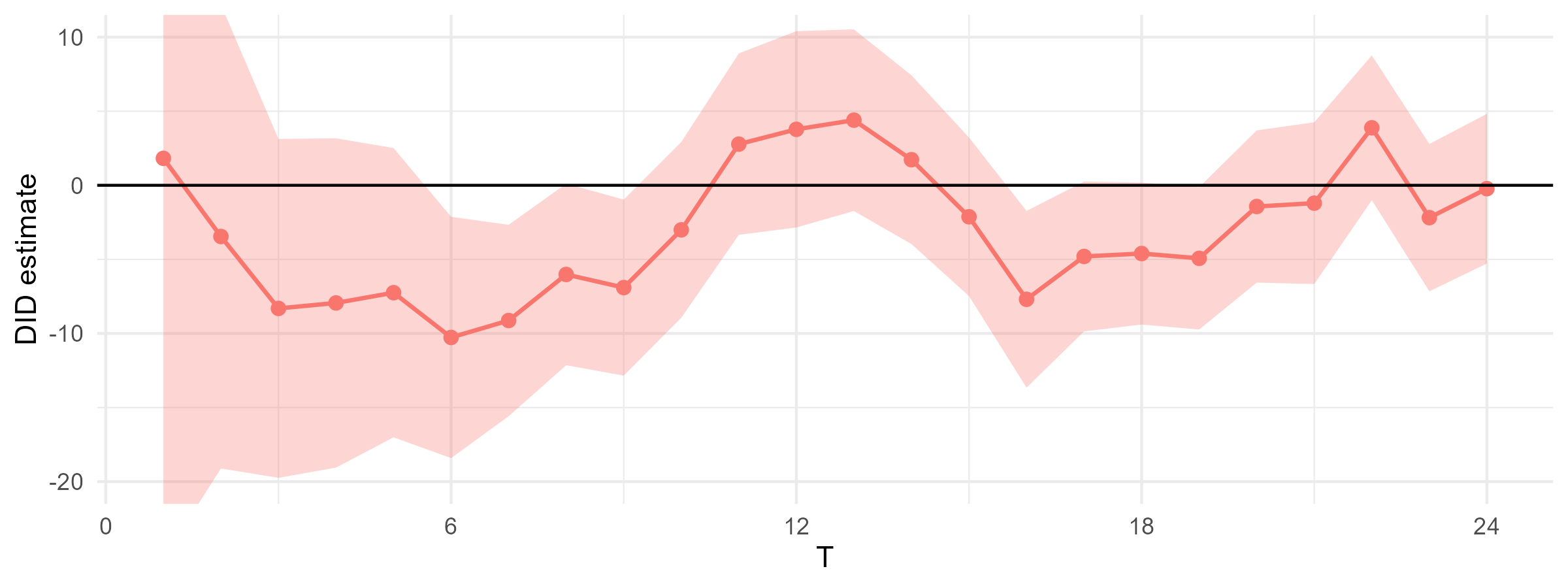}
        \caption{Edits, $\alpha=0.5$}
    \end{subfigure}
    \caption{Estimated DiD coefficients with various smoothing factors $\alpha$.
    }
    \label{fig:appx_expsmooth}
\end{figure}

\subsection{Variations in Binary Classification of Articles}
In our model described in Section~\ref{sec:models}, we classify each article as ``similar'' or ``dissimilar'' by first obtaining the set of articles $\mathcal{A}(T)$ that were created no more than $T$ months ago when their activities (views and edits) were observed, and then dividing $\mathcal{A}(T)$ evenly such that articles with similarity scores above the median in the set are considered similar. In this section, we present two variations of this methodology.

In Figure~\ref{fig:appx_global_sim_classifications}, we apply similarity classification on the entire pool of articles $\mathcal{A}(24)$ that are used in at least one of the diff-in-diff regressions. Each article then retains its fixed similarity label for all regressions with varying recency windows $T$. This means the individual regressions across the $x$-axis may not have the same number of similar vs. dissimilar articles.

\begin{figure}[h]
    \centering
    \begin{subfigure}[t]{0.4\textwidth}
        \centering
        \includegraphics[width=\linewidth]{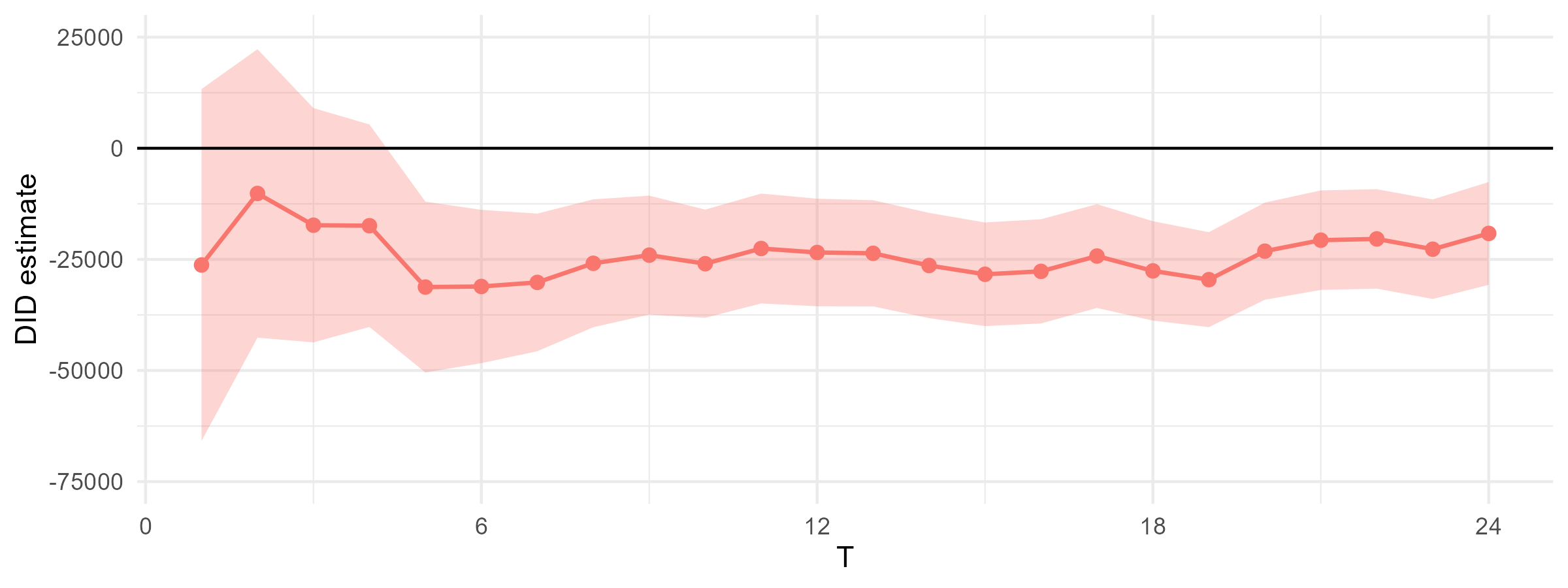}
        \caption{Views}
    \end{subfigure}
    \hfill
    \begin{subfigure}[t]{0.4\textwidth}
        \centering
        \includegraphics[width=\linewidth]{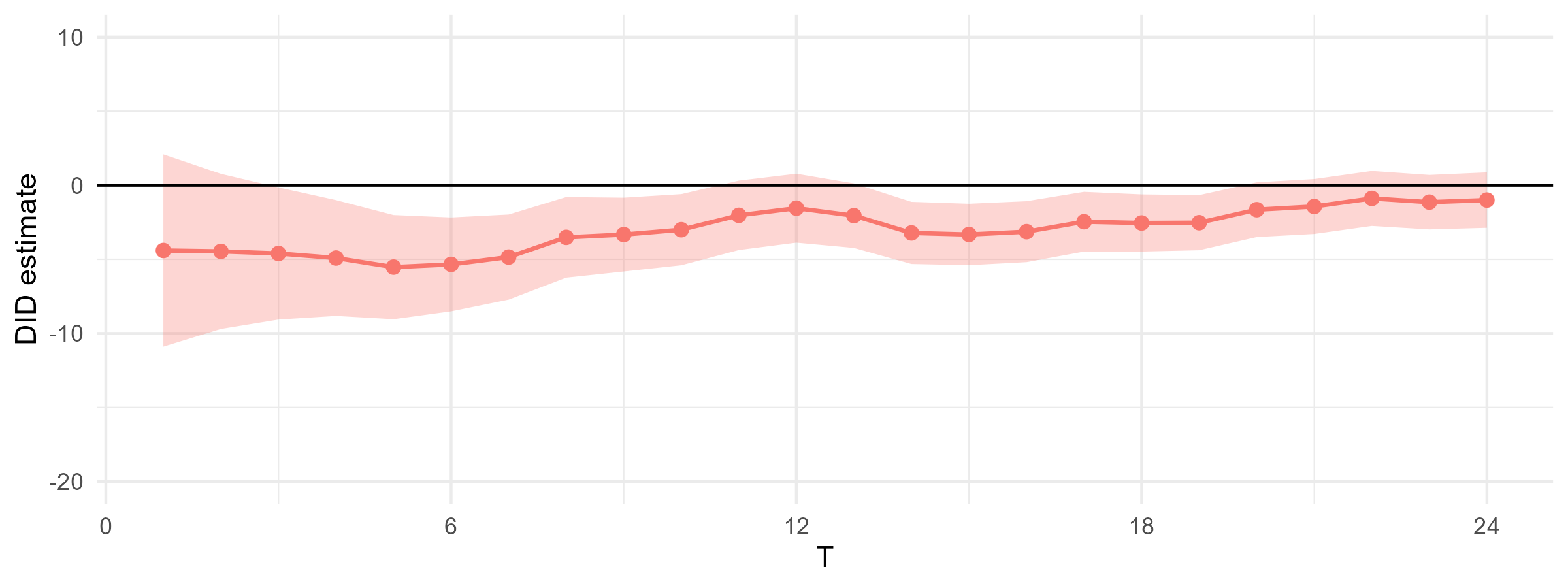}
        \caption{Edits}
    \end{subfigure}
    \caption{
    Estimated DiD coefficients when article classification is based on the entire pool $\mathcal{A}(24)$ and fixed for all regressions.
    }
    \label{fig:appx_global_sim_classifications}
\end{figure}

Alternatively, in Figures~\ref{fig:appx_remove_median} and \ref{fig:appx_remove_median_ts}, we still determine the classification labels based on individual pools of articles $\mathcal{A}(T)$ for each $T$; however, we remove articles whose similarity scores are near the median of this pool, in order to eliminate possible noise. In particular, if a fraction $\beta\in [0,1]$ of articles are removed, articles whose similarity scores are below the $((1-\beta)/2)$-quantile are considered dissimilar, and articles above the $(1-(1-\beta)/2)$-quantile are considered similar.

\begin{figure}[h]
    \centering
    \begin{subfigure}[t]{0.22\textwidth}
        \centering
        \includegraphics[width=\linewidth]{plots_appendix/Jan_9.1_DiD_exp_smooth_alpha_remove_median/exp=0.8_regression_Views_per_regression_0.50_1_sets_202112_to_202311.png}
        \caption{Views, $\beta=0$ (no removal)}
    \end{subfigure}
    \hfill
    \begin{subfigure}[t]{0.22\textwidth}
        \centering
        \includegraphics[width=\linewidth]{plots_appendix/Jan_9.1_DiD_exp_smooth_alpha_remove_median/exp=0.8_regression_Edits_per_regression_0.50_1_sets_202112_to_202311.png}
        \caption{Edits, $\beta=0$ (no removal)}
    \end{subfigure}
    \hfill
    \begin{subfigure}[t]{0.22\textwidth}
        \centering
        \includegraphics[width=\linewidth]{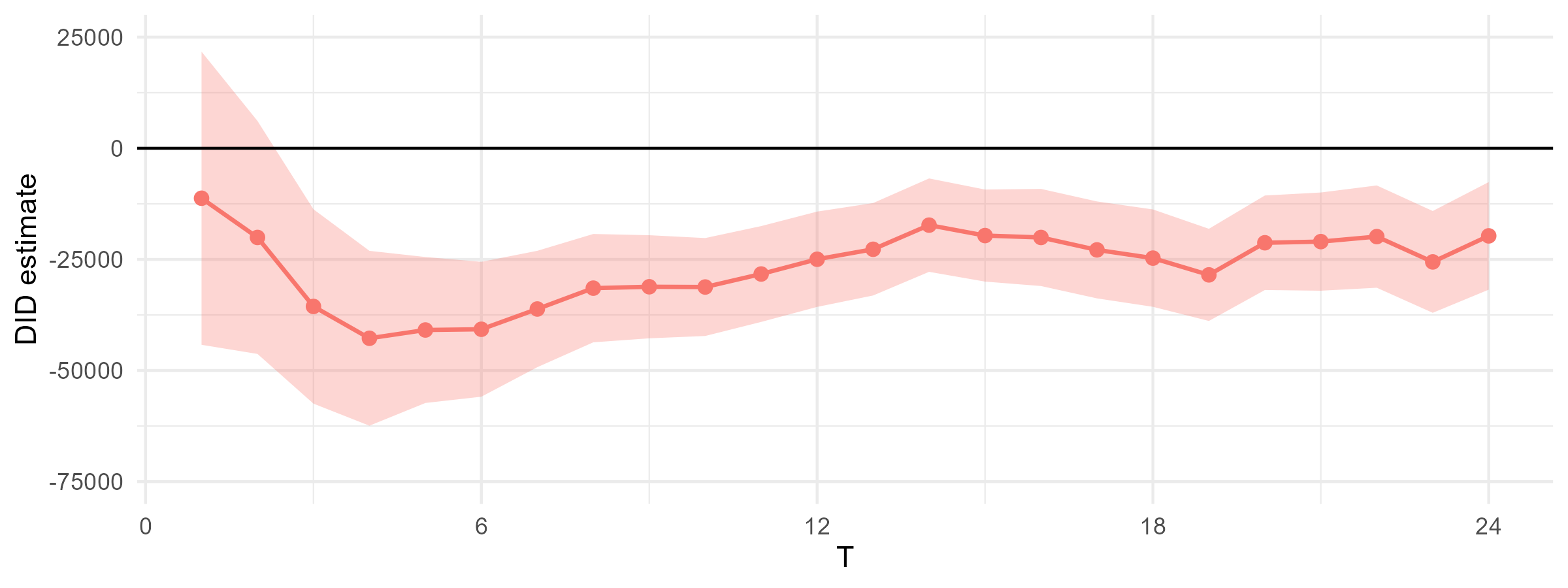}
        \caption{Views, $\beta=0.1$}
    \end{subfigure}
    \hfill
    \begin{subfigure}[t]{0.22\textwidth}
        \centering
        \includegraphics[width=\linewidth]{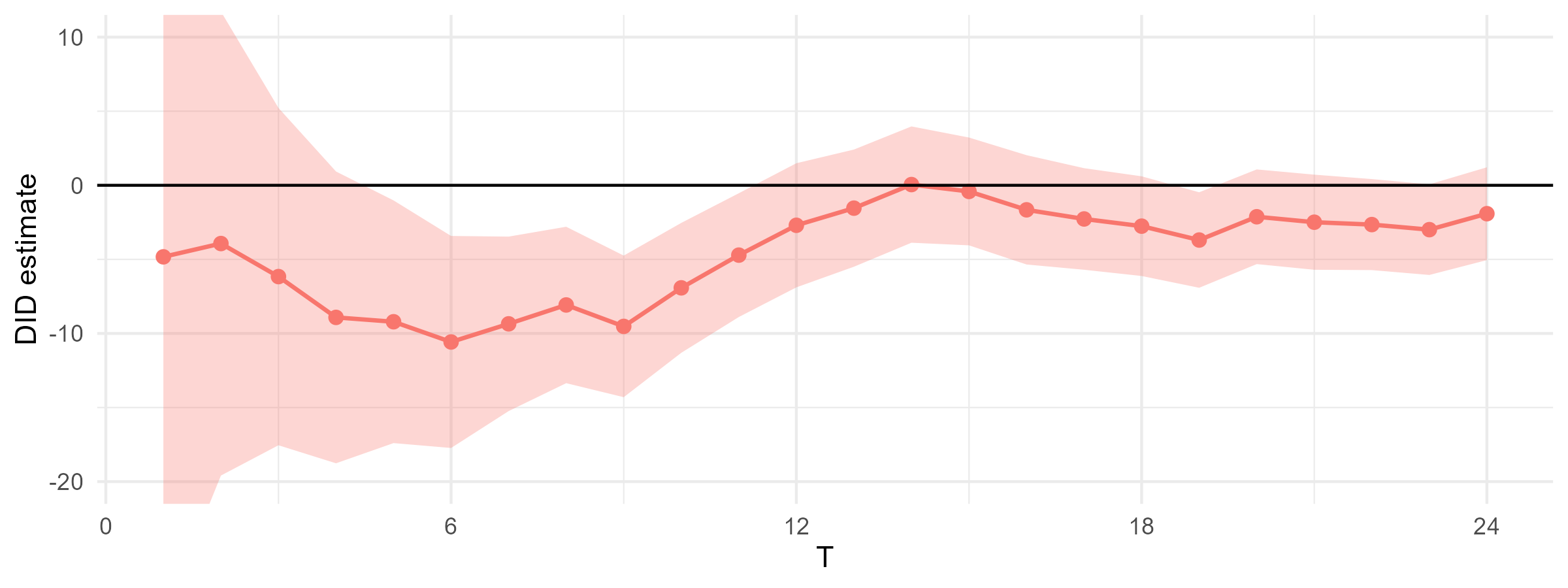}
        \caption{Edits, $\beta=0.1$}
    \end{subfigure}
    \hfill
    \begin{subfigure}[t]{0.22\textwidth}
        \centering
        \includegraphics[width=\linewidth]{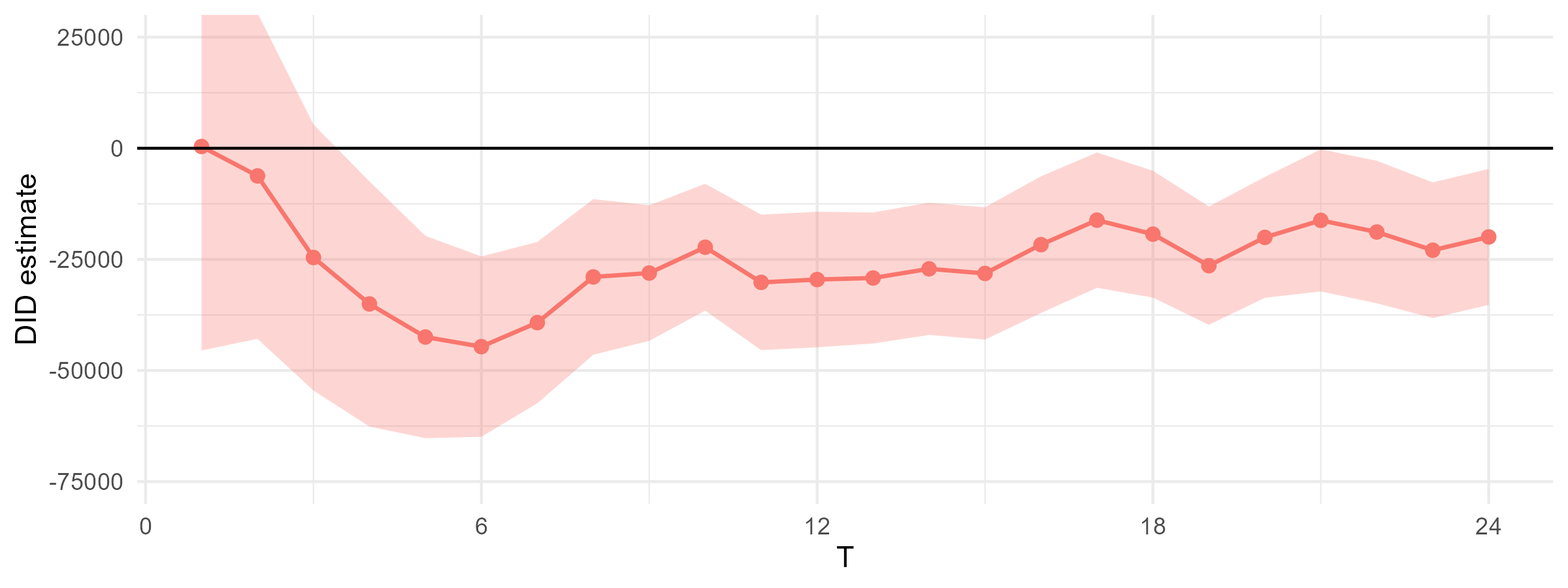}
        \caption{Views, $\beta=0.5$}
    \end{subfigure}
    \hfill
    \begin{subfigure}[t]{0.22\textwidth}
        \centering
        \includegraphics[width=\linewidth]{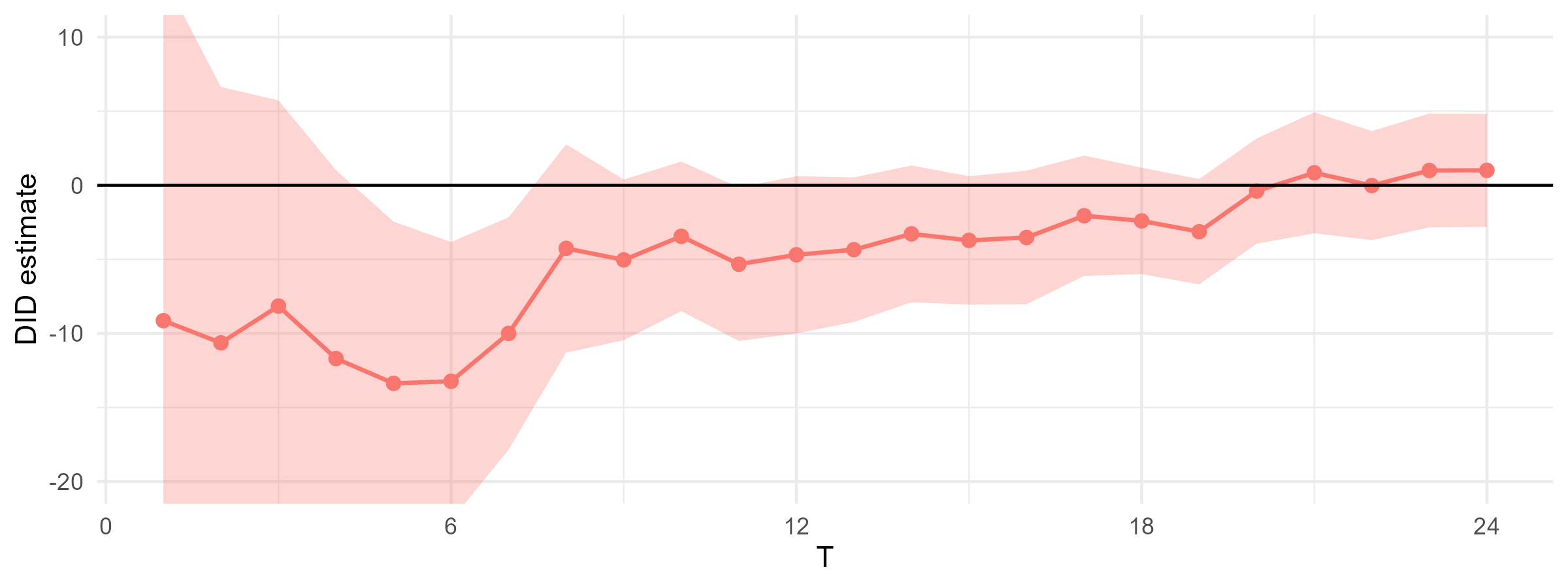}
        \caption{Edits, $\beta=0.5$}
    \end{subfigure}
    \caption{Estimated DiD coefficients after removing $\beta$ fraction of articles with similarity scores near the median.
    }
    \label{fig:appx_remove_median}
\end{figure}

\begin{figure}[h]
    \centering
    \begin{subfigure}[t]{0.22\textwidth}
        \centering
        \includegraphics[width=\linewidth]{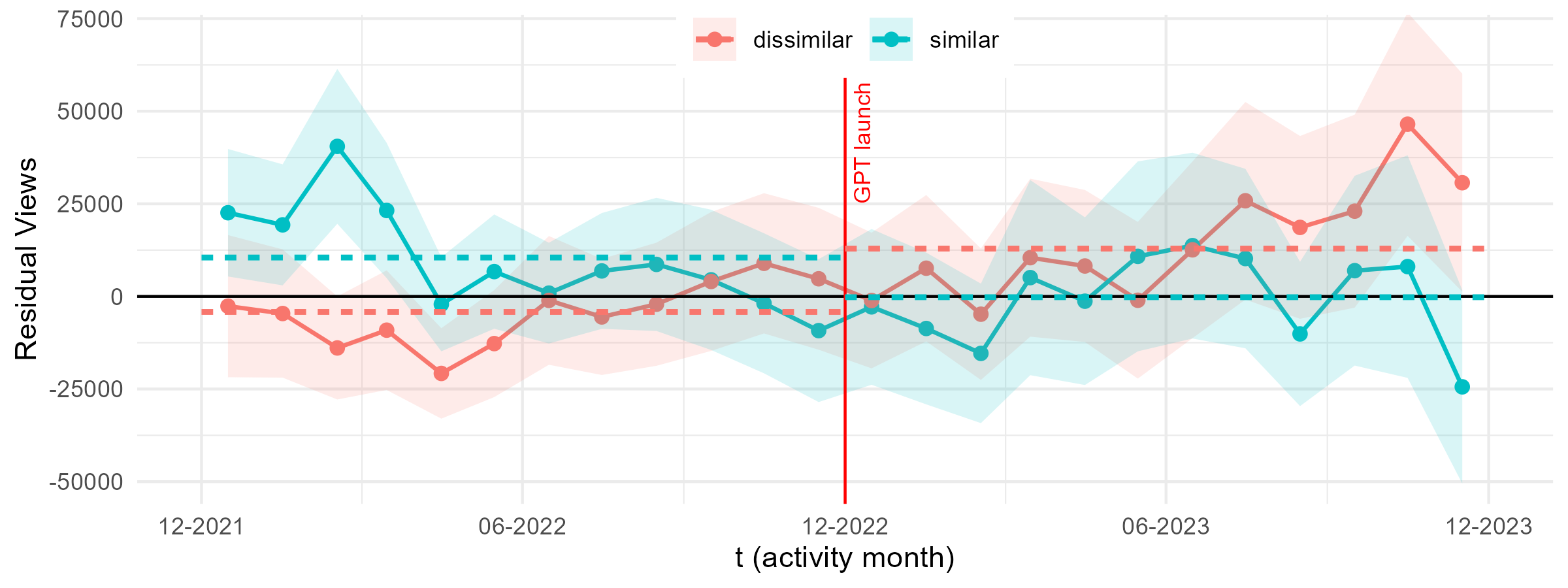}
        \caption{Views, $\beta=0$ (no removal)}
    \end{subfigure}
    \hfill
    \begin{subfigure}[t]{0.22\textwidth}
        \centering
        \includegraphics[width=\linewidth]{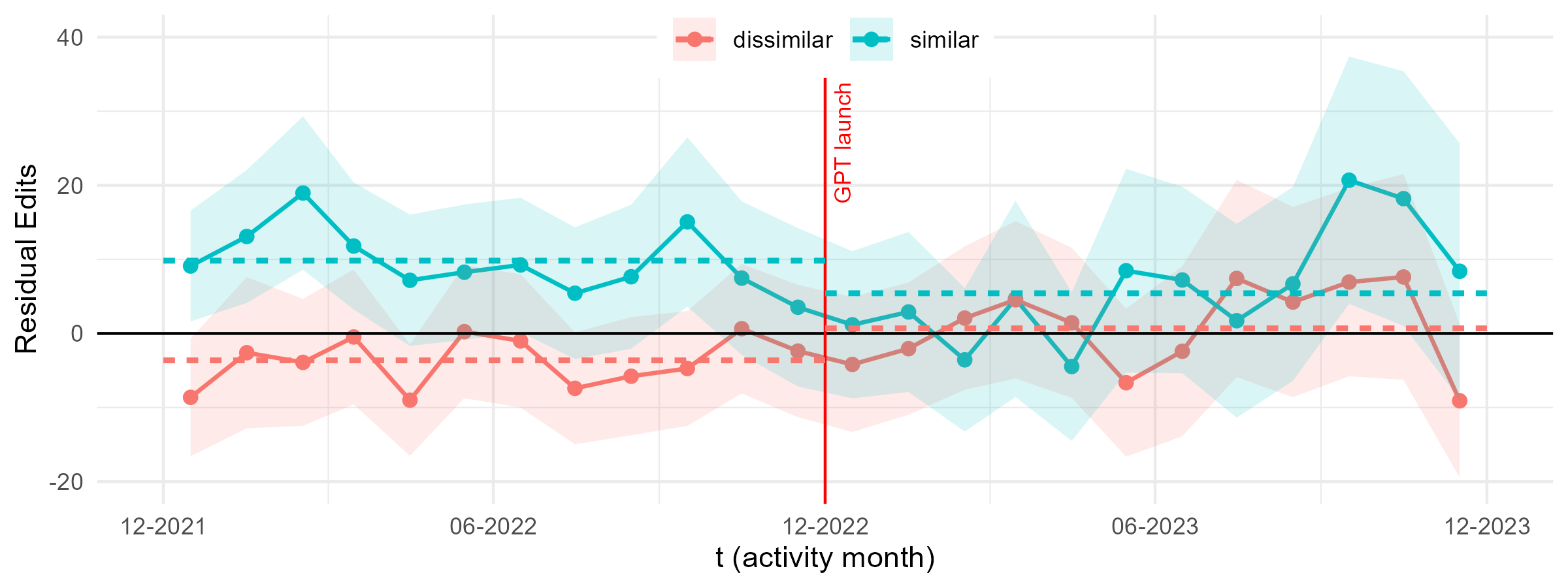}
        \caption{Edits, $\beta=0$ (no removal)}
    \end{subfigure}
    \hfill
    \begin{subfigure}[t]{0.22\textwidth}
        \centering
        \includegraphics[width=\linewidth]{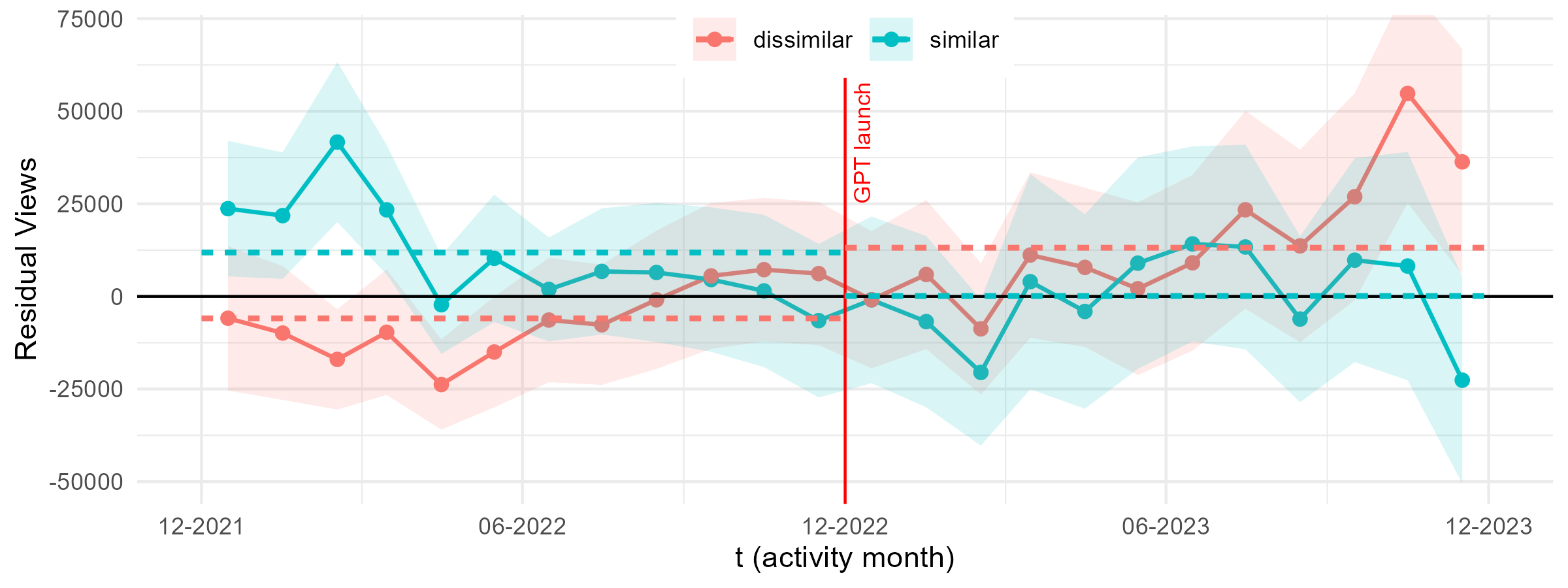}
        \caption{Views, $\beta=0.1$}
    \end{subfigure}
    \hfill
    \begin{subfigure}[t]{0.22\textwidth}
        \centering
        \includegraphics[width=\linewidth]{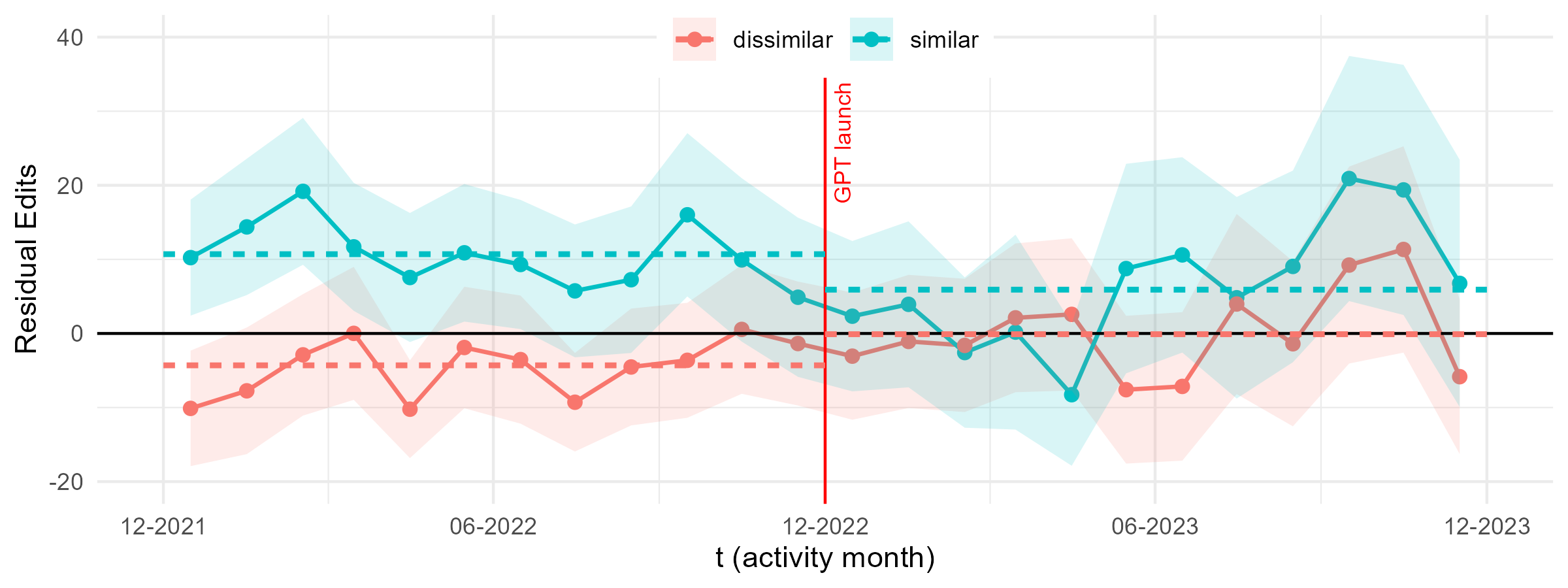}
        \caption{Edits, $\beta=0.1$}
    \end{subfigure}
    \hfill
    \begin{subfigure}[t]{0.22\textwidth}
        \centering
        \includegraphics[width=\linewidth]{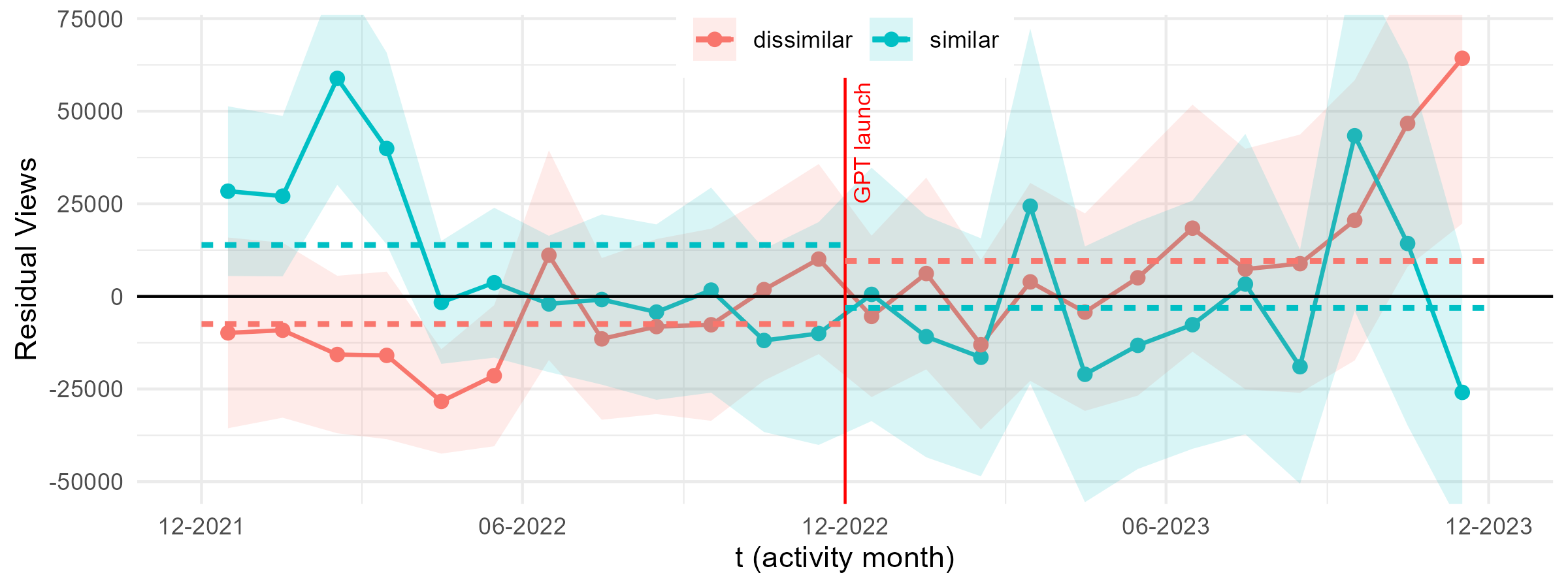}
        \caption{Views, $\beta=0.5$}
    \end{subfigure}
    \hfill
    \begin{subfigure}[t]{0.22\textwidth}
        \centering
        \includegraphics[width=\linewidth]{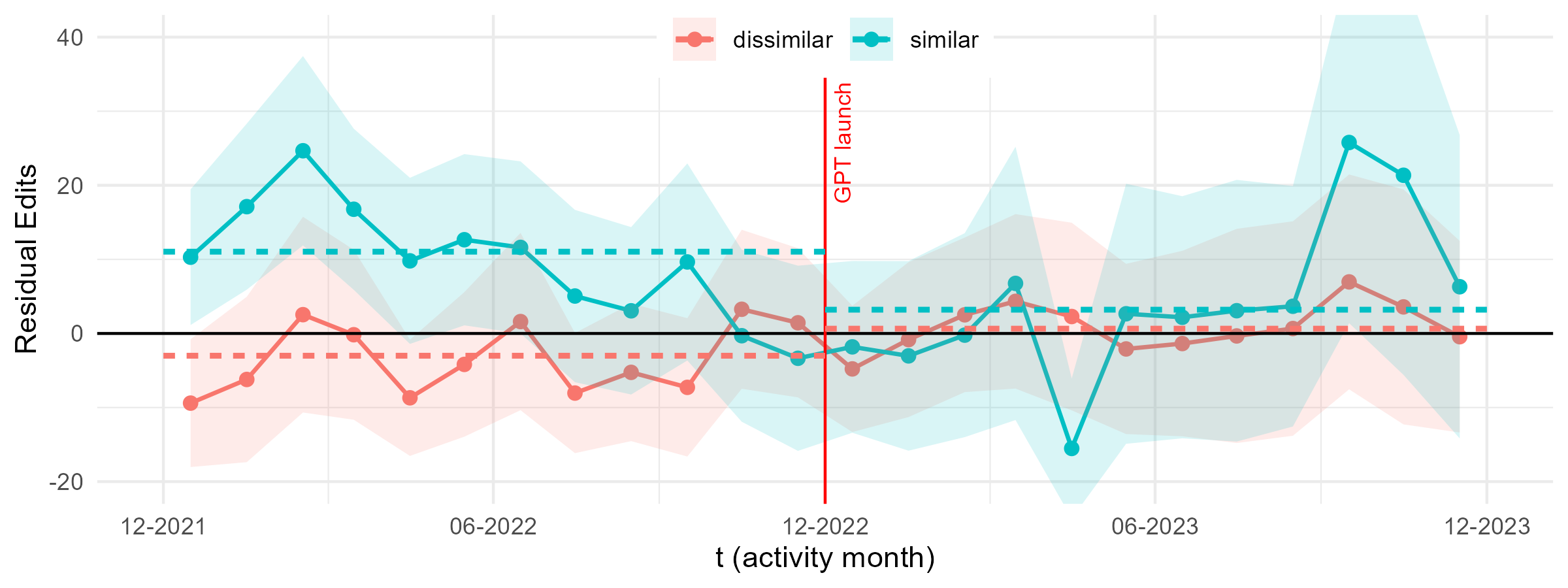}
        \caption{Edits, $\beta=0.5$}
    \end{subfigure}
    \caption{Mean residuals for each month $t$ in which activity occurs among all articles in $\mathcal{A}(6)$, after removing $\beta$ fraction of articles with similarity scores near the median.
    }
    \label{fig:appx_remove_median_ts}
\end{figure}

\section{Falsification Tests}
In this section, we present alternative datasets that do not show the same effects as the main text, confirming the validity of our analysis. We also provide possible explanations for why no such effects are observed in each of these falsification tests.

\subsection{Random Articles}
In the main analysis, we use Wikipedia articles that were among the 1000 pages with the most views sitewide in any given month of the event study period. In contrast, this section uses 10000 randomly sampled articles on English Wikipedia. 

\begin{figure}[h]
    \centering
    \begin{subfigure}[t]{0.48\textwidth}
        \centering
        \includegraphics[width=\linewidth]{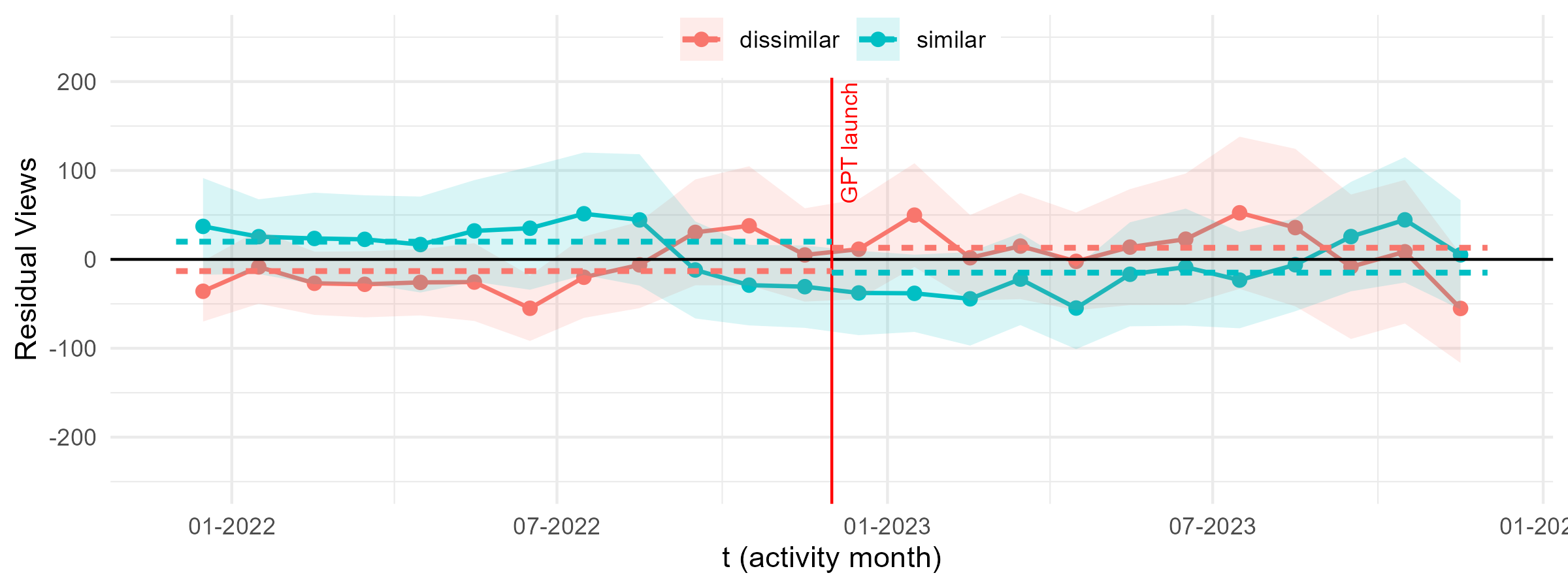}
        \caption{Views}
    \end{subfigure}
    \hfill
    \begin{subfigure}[t]{0.48\textwidth}
        \centering
        \includegraphics[width=\linewidth]{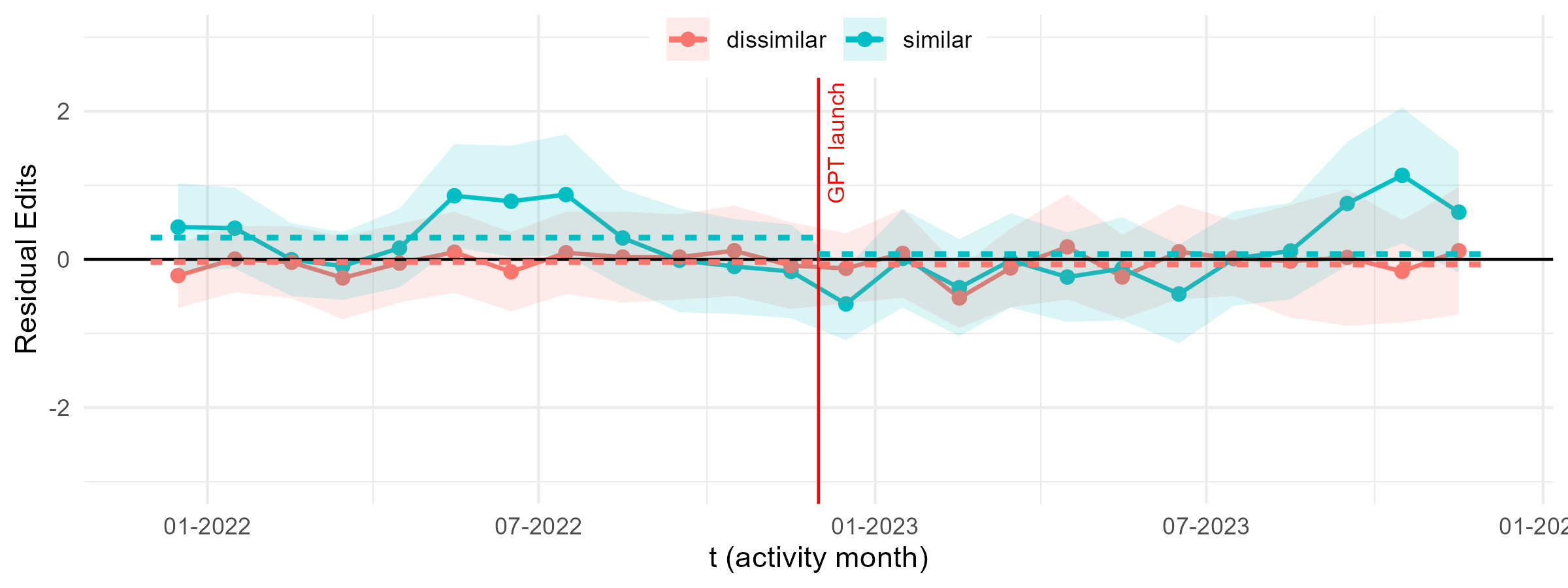}
        \caption{Edits}
    \end{subfigure}
    \caption{Time series of mean residuals for 10000 random articles. (Note that the $y$-axis is different in scale compared to Figure~\ref{fig:time_series_popularnew_single_reg_bootstrap}.)
    }
    \label{fig:appx_false_random_ts}
\end{figure}

\begin{figure}[h]
    \centering
    \begin{subfigure}[t]{0.48\textwidth}
        \centering
        \includegraphics[width=\linewidth]{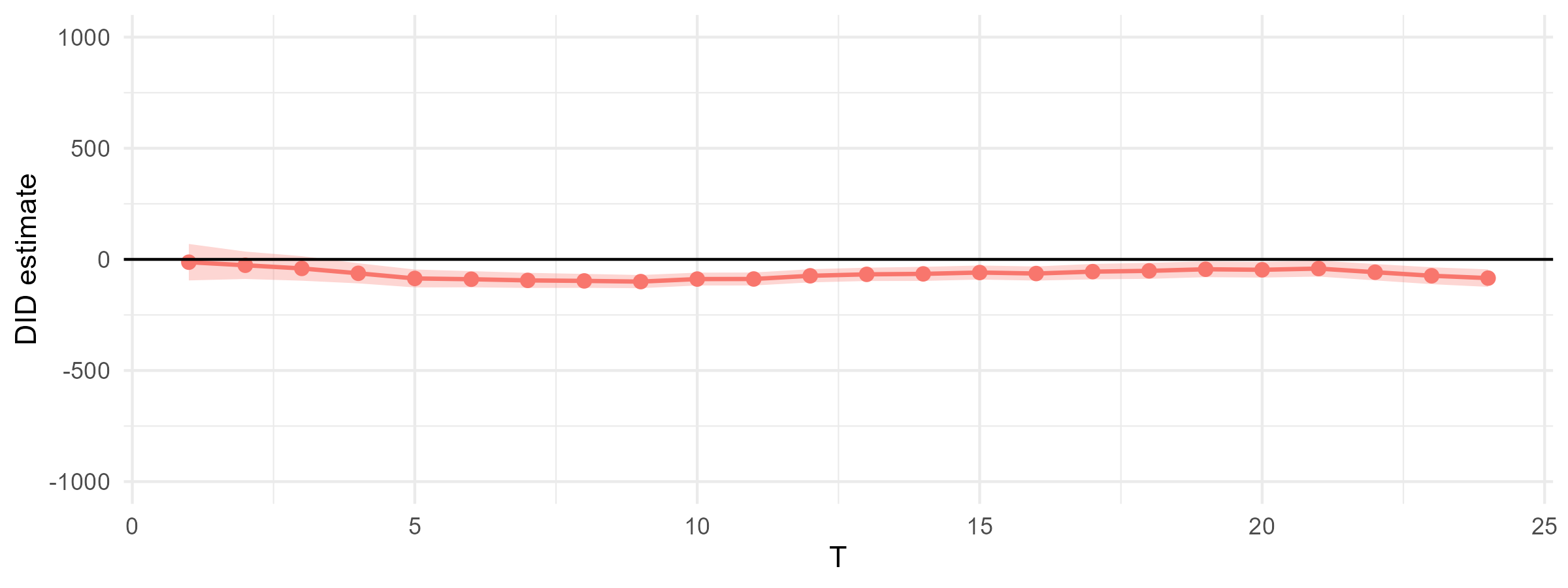}
        \caption{Views}
    \end{subfigure}
    \hfill
    \begin{subfigure}[t]{0.48\textwidth}
        \centering
        \includegraphics[width=\linewidth]{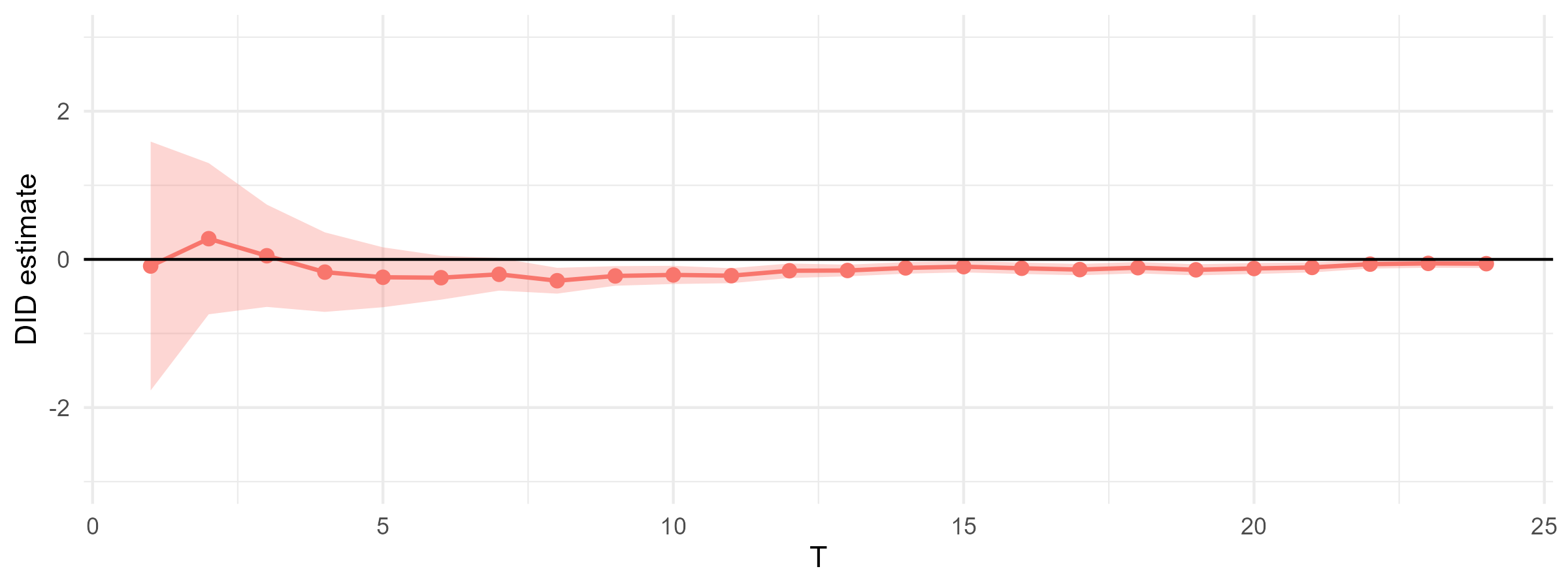}
        \caption{Edits}
    \end{subfigure}
    \caption{Estimated DiD coefficients for 10000 random articles. (Note that the $y$-axis is different in scale compared to Figure~\ref{fig:did_3.5small_popularnewA}.)
    }
    \label{fig:appx_false_random}
\end{figure}

Figures~\ref{fig:appx_false_random_ts} and \ref{fig:appx_false_random} show that, while there may be some hints of effects on random articles similar to those on popular articles, these effects are much less pronounced, less conclusive, and often statistically insignificant.
We hypothesize that the lack of conclusive results from random articles may be due to differences in how users interact with them. Because the vast majority of Wikipedia articles are inactive and concern niche topics, users who view and edit them are more likely to do so out of curiosity or self-exploration, and thus they may be more commited to Wikipedia over other means of information. In contrast, users who view popular articles on Wikipedia may be simply in need of a factual lookup, making ChatGPT a more plausible alternative.

\subsection{GPT 4o-mini}  \label{sec:appx_gpt4}
We used GPT 3.5 Turbo to generate the LLM articles, as explained in Section~\ref{sec:gpt_sim}. In this section, we instead use GPT 4o-mini, the latest ChatGPT model that is fully accessible to free users as of writing, to generate the responses.

As seen from Figures~\ref{fig:appx_false_gpt4_ts} and \ref{fig:appx_false_gpt4}, we no longer observe the key results from our main study: namely, there is no longer any significant difference between the growth in activity between similar and dissimilar articles. This may be because GPT 4o-mini's training data includes most articles in our dataset, so the model likely performs better on their subject matters than GPT 3.5 Turbo does. Not only would this make GPT 4o-mini a better substitute for Wikipedia, but it also reduces the variance in substitutability among different articles, so the semantic similarity between true Wikipedia articles and LLM responses may no longer be as effective in measuring user preferences of the two platforms.

\begin{figure}[h!]
    \centering
    \begin{subfigure}[t]{0.48\textwidth}
        \centering
        \includegraphics[width=\linewidth]{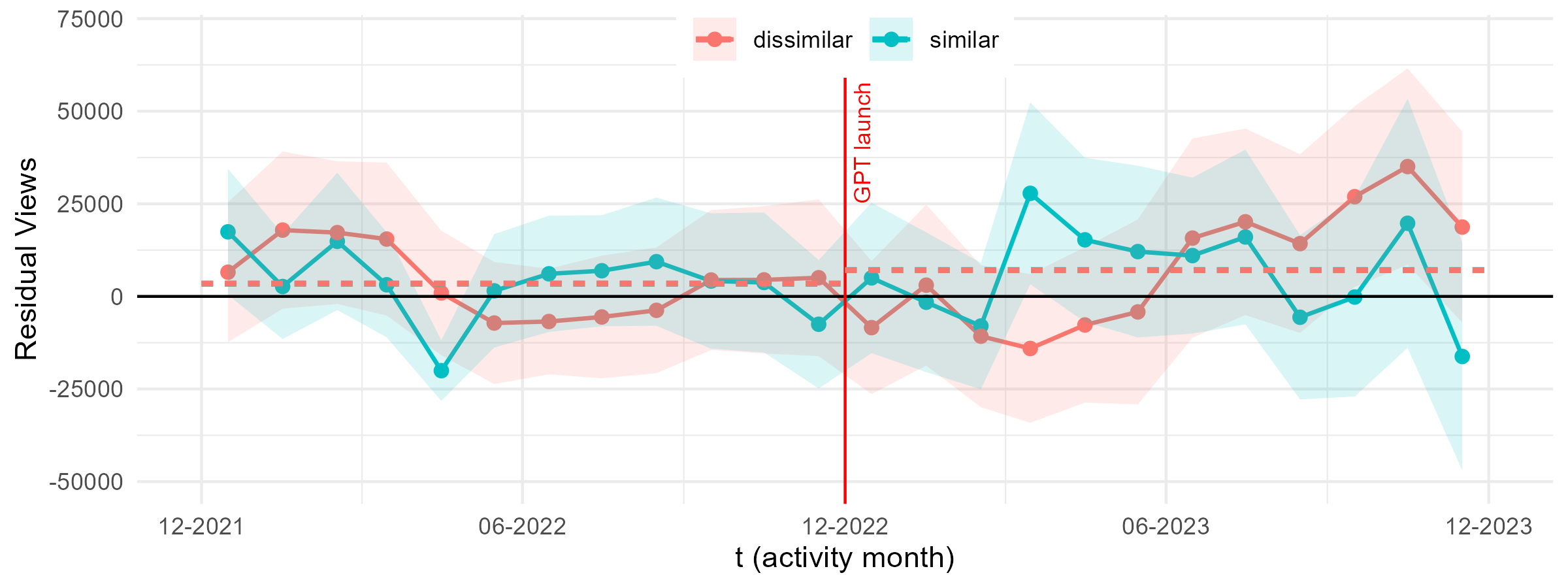}
        \caption{Views}
    \end{subfigure}
    \hfill
    \begin{subfigure}[t]{0.48\textwidth}
        \centering
        \includegraphics[width=\linewidth]{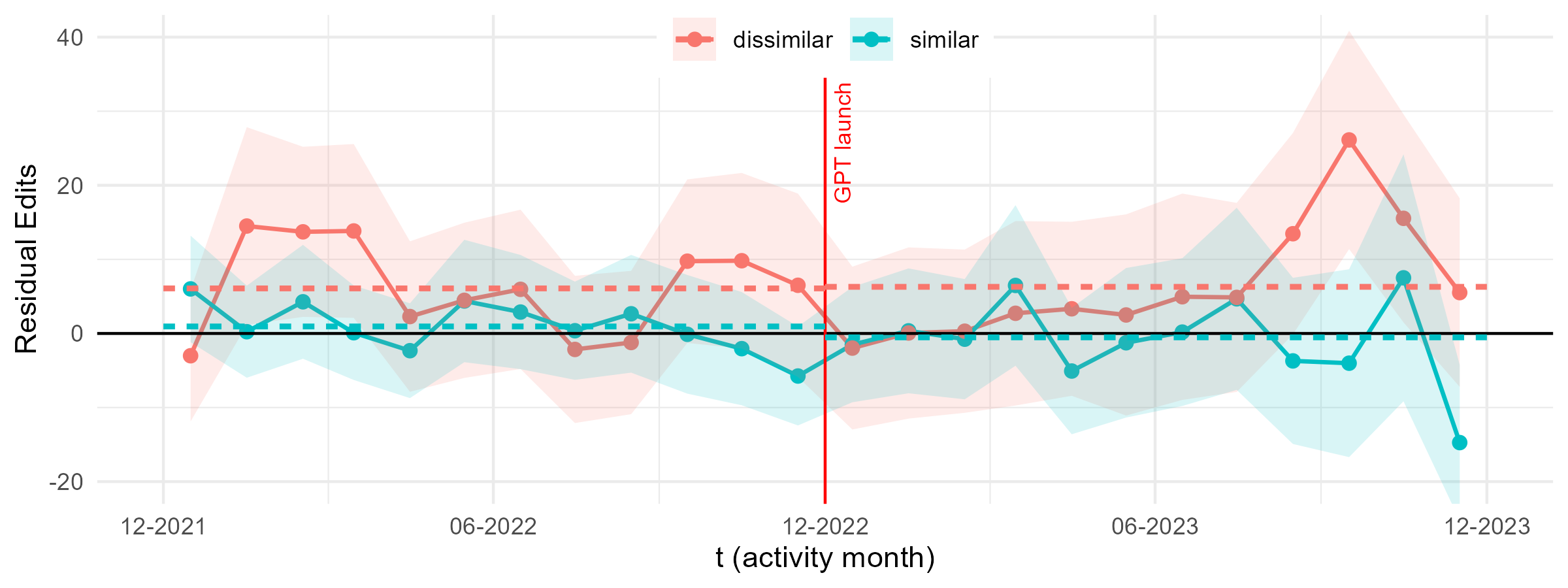}
        \caption{Edits}
    \end{subfigure}
    \caption{Time series of mean residuals with LLM responses generated by GPT 4o-mini
    }
    \label{fig:appx_false_gpt4_ts}
\end{figure}

\begin{figure}[h!]
    \centering
    \begin{subfigure}[t]{0.48\textwidth}
        \centering
        \includegraphics[width=\linewidth]{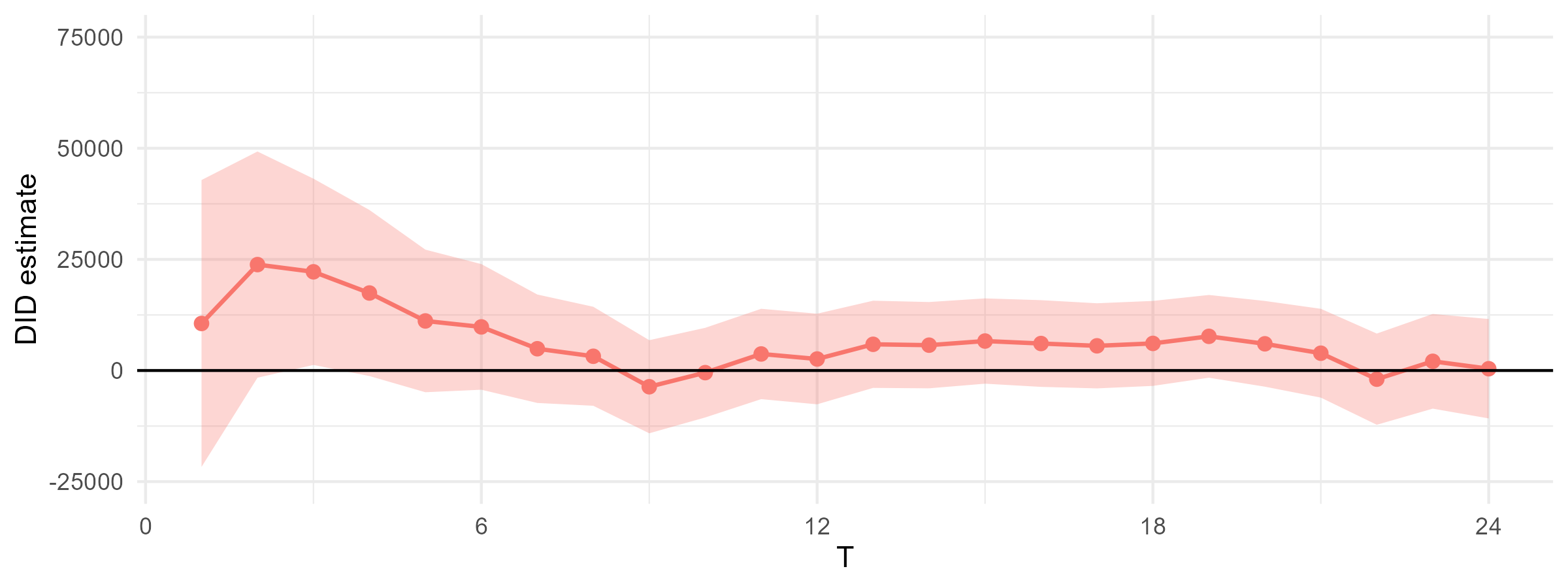}
        \caption{Views}
    \end{subfigure}
    \hfill
    \begin{subfigure}[t]{0.48\textwidth}
        \centering
        \includegraphics[width=\linewidth]{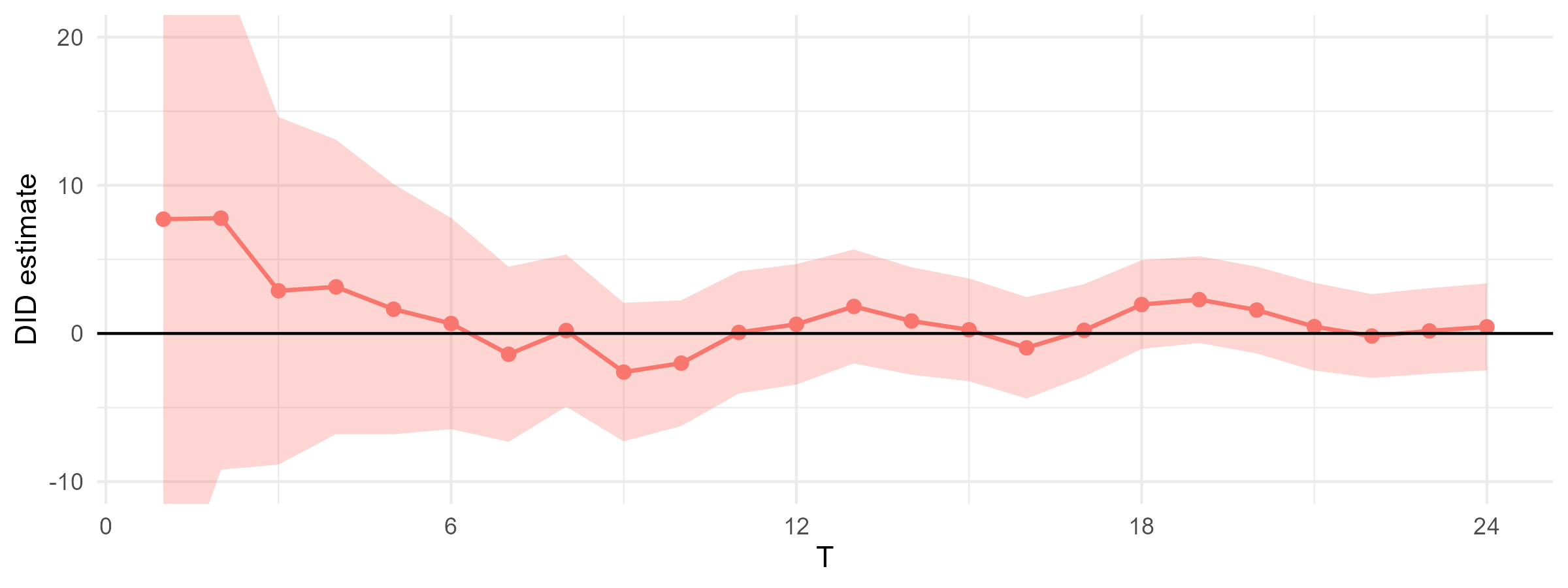}
        \caption{Edits}
    \end{subfigure}
    \caption{Estimated DiD coefficients with LLM responses generated by GPT 4o-mini
    }
    \label{fig:appx_false_gpt4}
\end{figure}

\subsection{Generation Date of GPT Responses}
While our main analysis uses GPT 3.5 responses generated between April and June 2024, Figures~\ref{fig:appx_false_dec24_ts} and \ref{fig:appx_false_dec24} instead use responses generated in December 2024 with the same model.
The results for views are still consistent with the earlier results, but not edits. This supports our main findings of significant effects on views, and suggestive but inconclusive effects on edits.

We remark that the differences here may also be due to model training, similarly to the previous section. Even though GPT 3.5 Turbo was a legacy model as of December 2024, it may still have been retrained or fine-tuned during this period. As such, it may have picked up further information from articles in the dataset, similar to GPT 4o-mini.

\begin{figure}[h!]
    \centering
    \begin{subfigure}[t]{0.48\textwidth}
        \centering
        \includegraphics[width=\linewidth]{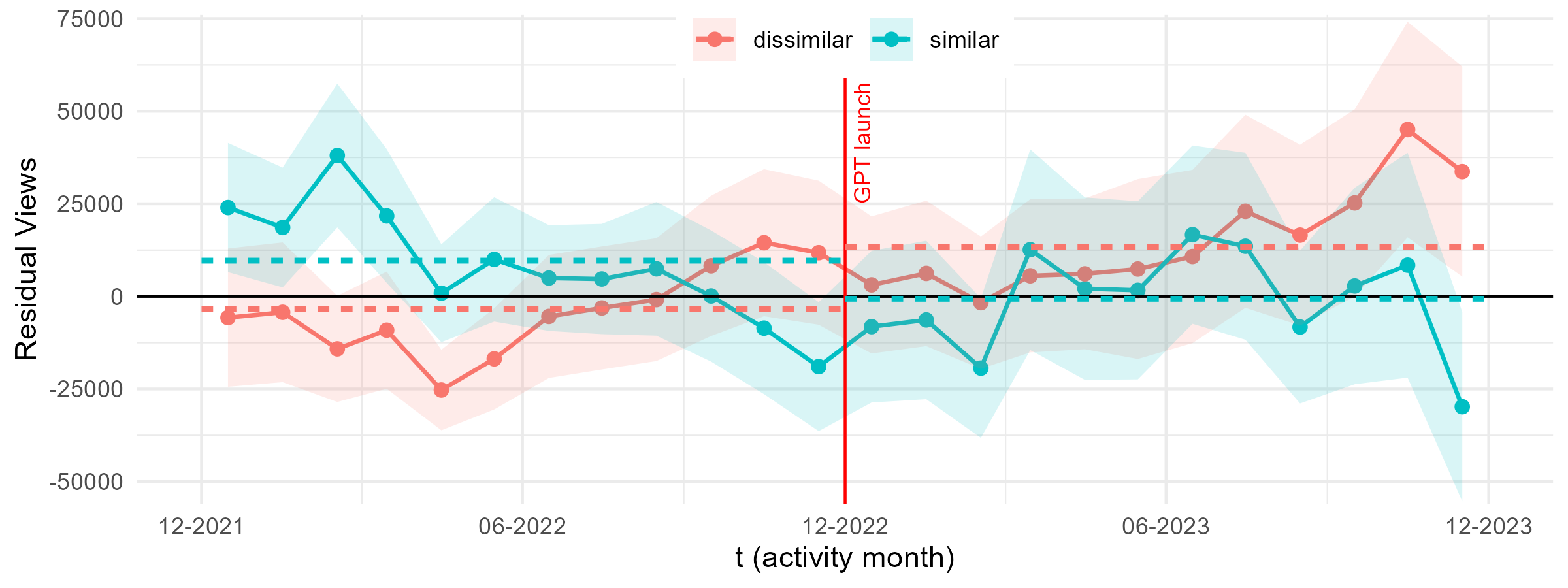}
        \caption{Views}
    \end{subfigure}
    \hfill
    \begin{subfigure}[t]{0.48\textwidth}
        \centering
        \includegraphics[width=\linewidth]{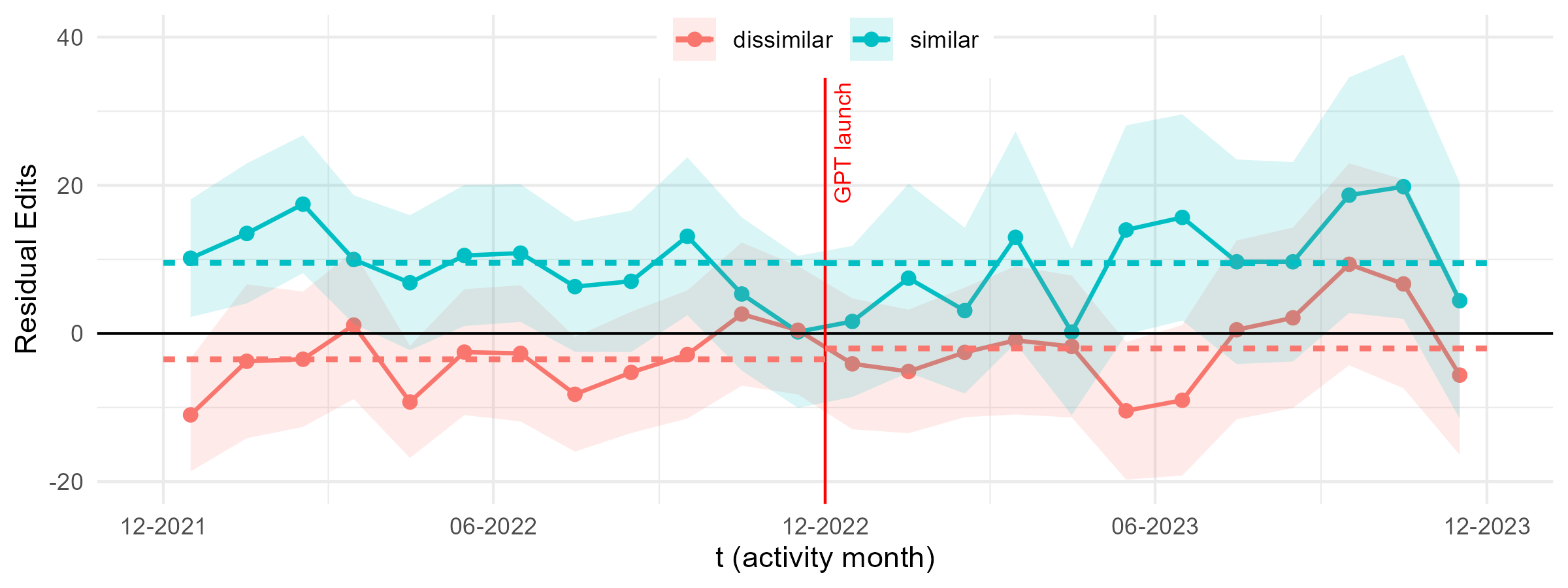}
        \caption{Edits}
    \end{subfigure}
    \caption{Time series of mean residuals with LLM responses generated by GPT 4o-mini
    }
    \label{fig:appx_false_dec24_ts}
\end{figure}

\begin{figure}[h!]
    \centering
    \begin{subfigure}[t]{0.48\textwidth}
        \centering
        \includegraphics[width=\linewidth]{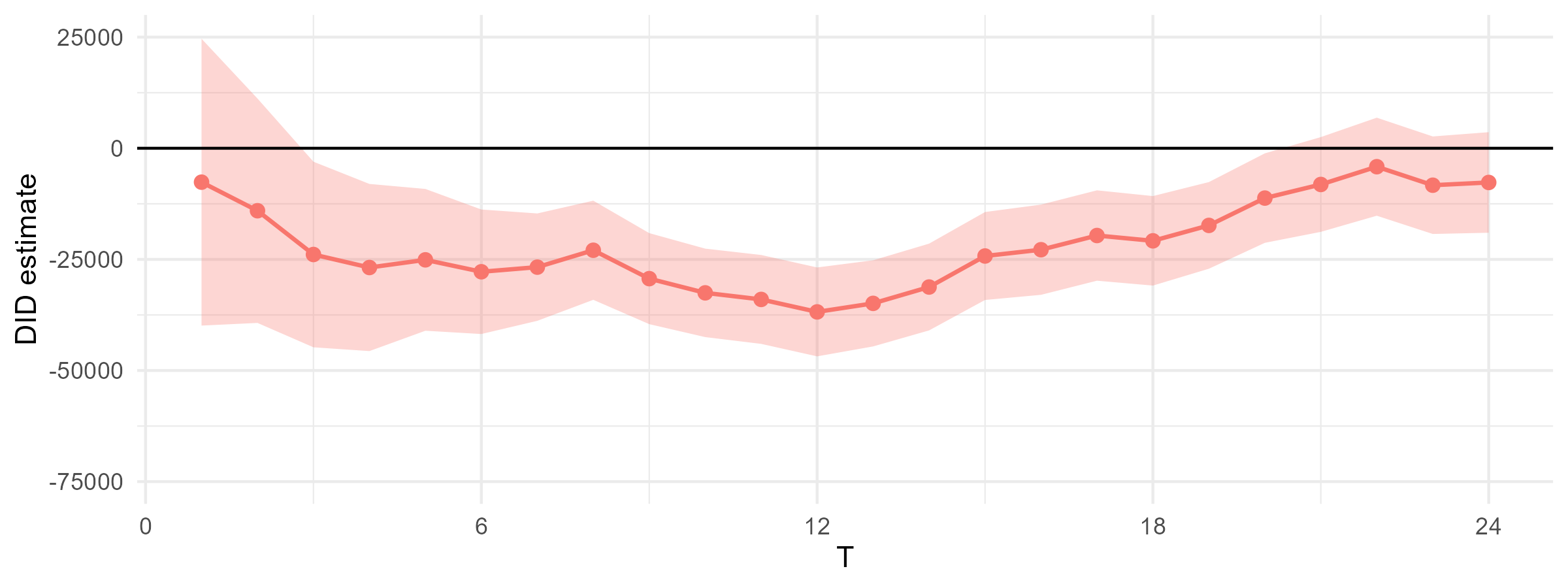}
        \caption{Views}
    \end{subfigure}
    \hfill
    \begin{subfigure}[t]{0.48\textwidth}
        \centering
        \includegraphics[width=\linewidth]{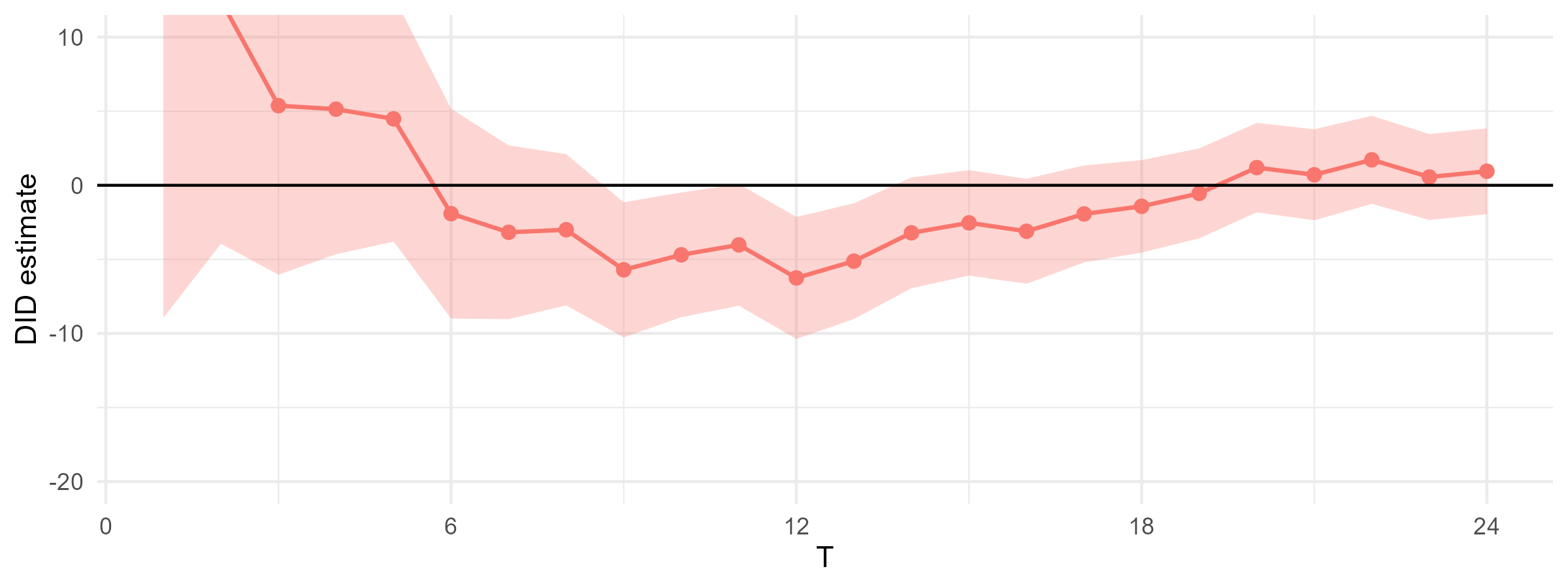}
        \caption{Edits}
    \end{subfigure}
    \caption{Estimated DiD coefficients with LLM responses generated by GPT 4o-mini
    }
    \label{fig:appx_false_dec24}
\end{figure}

\fi

\end{document}